\documentclass[fleqn,usenatbib]{mnras}

\usepackage{newtxtext,newtxmath}

\usepackage[T1]{fontenc}

\DeclareRobustCommand{\VAN}[3]{#2}
\let\VANthebibliography\thebibliography
\def\thebibliography{\DeclareRobustCommand{\VAN}[3]{##3}\VANthebibliography}

\usepackage{graphicx}	
\usepackage{amsmath}	
\usepackage[dvipsnames]{xcolor}

\title[Magnetic frame-dragging in compact neutron stars]{Magnetic frame-dragging correction to the electromagnetic solution of a compact neutron star}

\author[R. Torres et al.]{
R. Torres,$^{1}$\thanks{E-mail: rui.t.torres@tecnico.ulisboa.pt}
T. Grismayer,$^{1}$
F. Cruz,$^{1,2}$
and L.O. Silva$^{1}$
\\
$^{1}$GoLP/Instituto de Plasmas e Fus\~{a}o Nuclear, Instituto Superior T\'{e}cnico, Universidade de Lisboa, 1049-001 Lisboa, Portugal\\
$^{2}$Inductiva Research Labs, Rua da Prata 80, 1100-420 Lisboa, Portugal
}

\date{Accepted XXX. Received YYY; in original form ZZZ}

\pubyear{2023}

\begin{document}
\label{firstpage}
\pagerange{\pageref{firstpage}--\pageref{lastpage}}
\maketitle

\begin{abstract}
{Neutron stars are usually modelled as spherical, rotating perfect conductors with a predominant intrinsic dipolar magnetic field anchored to their stellar crust. Due to their compactness, General Relativity corrections must be accounted for in Maxwell's equations, leading to modified interior and exterior electromagnetic solutions. We present analytical solutions for slowly-rotating magnetised neutron stars taking into account the magnetic frame-dragging correction. For typical compactness values, i.e. $R_s \sim 0.5 [R_*]$, we show that the new terms lead to a percent order correction in the magnetic field orientation and strength compared to the case with no magnetic frame-dragging correction. Also, we obtain a self-consistent redistribution of the surface azimuthal current. We verify the validity of the derived solution through two-dimensional particle-in-cell simulations of an isolated neutron star. Defining the azimuthal electric and magnetic field amplitudes during the transient phase as observables, we prove that the magnetic frame-dragging correction reduces the transient wave amplitude, as expected from the analytical solution. We show that simulations are more accurate and stable when we include all first-order terms. The increased accuracy at lower spatiotemporal resolutions translates into a reduction in simulation runtimes.}
\end{abstract}

\begin{keywords}
magnetic fields -- methods: analytical -- methods: numerical -- stars: general -- stars: neutron – pulsars: general
\end{keywords}



\section{Introduction}

Compact objects have long been theorised to power non-thermal emission. In their vicinity, general-relativistic (GR) effects can couple to strong electromagnetic (EM) fields and promote particle acceleration, pair creation and, potentially, pulsed emission. These plasma-mediated processes are dependent on the underlying magnetic field configuration. In neutron stars, these effects are critically important to take into account due to their very high intrinsic magnetic fields, of order {$10^{8}-10^{14}$ G \citep[e.g.][]{Igoshev2021}}.

The stationary electromagnetic solution to Maxwell's equations of an idealised magnetic star was first obtained by \citet{Deutsch1955}. This solution considered a perfect rotating spherical conductor with a misalignment between the dipolar magnetic moment and rotation axis, in flat spacetime background metric (Minkowski). The first works to include the effects of curved spacetimes were for non-rotating neutron stars, i.e. using a Schwarzschild background metric. These solutions have shown that the amplitude of the magnetic field increases in comparison to the flat spacetime case when considering the same dipolar moment \citep{Ginzburg1964,Wasserman1983,Petterson1974} or multipolar moments \citep{Anderson1970}. As for slow-rotating neutron stars, the rigid-body rotation of the compact mass leads to a general-relativistic effect called the drag of the inertial frames or frame-dragging effect. This effect induces an electric field close to the central body that was shown to decrease the magnitude of the unipolar induction from the rotation of the stellar crust \citep{Muslimov1986}. Solutions for the exterior electromagnetic fields were obtained for the aligned rotator \citep{Konno2000} and the oblique rotator configuration \citep{Rezzolla2001a,Rezzolla2001b}. These works explored the implications for leptonic acceleration and subsequent radiation in pulsar vacuum gaps. Another approach allowed the exterior field lines to move faster than the crust, e.g. for a fast-rotating neutron star core \citep{Muslimov1986,Kojima2004}. More recently, several numerical works, employing spectral methods to solve Maxwell's equations in the slow-rotation approximation of general relativity, have obtained approximate solutions up to third order in the spin parameter \citep{Petri2013}. This work was generalised for the multipolar case in the strong gravity regime \citep{Petri2017} to show that introducing small-scale magnetic field structures could enhance pair production in the vacuum gap and increase the plasma multiplicity, in agreement with observations.

The purpose of this paper is to extend the set of analytical solutions available in the literature to include the frame-dragging effect in both the electric and magnetic fields for the aligned rotator configuration. As we will show here, these corrections lead to more accurate simulations, with potential savings in runtimes. We use the 3+1 formalism of electrodynamics in curved spacetime as described in Section~\ref{section:2}. In Section~\ref{section:3}, we show how to obtain the stationary solution to an aligned rotator. In Section~\ref{section:4}, we introduce the numerical setup that is used in Section~\ref{section:5} for a numerical realization of the solution derived, using a general-relativistic particle-in-cell code. Conclusions and future prospects are outlined in Section~\ref{section:6}.

\section{General Relativity}
\label{section:2}

{The electromagnetic energy density of highly magnetised stars is orders of magnitude smaller than the corresponding total mass density. Hence, we can avoid the coupled Einstein-Maxwell's equations and solve the general-relativistic Maxwell's equations on a fixed curved background metric.}

{Throughout the paper, we use units in which $c=G=1$, quantities subscripted with an asterisk denote stellar properties (e.g., $R_*$ for the stellar radius) and ($-$,$+$,$+$,$+$) metric signature. In addition, we adopt the 3+1 formalism of general relativity \citep{Thorne1982}.}

\subsection{Slowly-rotating spacetime}

{The seminal papers by \citet{Hartle1967,Hartle1968} described the background metric of a rotating compact neutron star which assumes a one-parameter equation of state, axial symmetry, and slow-uniform rotation. The last condition states that the stellar angular velocity must be much smaller than the light speed ($R_*\Omega_*\ll1$). In this approximation, and keeping only terms up to first order in the angular velocity, the background metric takes the form \citep{Hartle1967}:}
\begin{equation}
    \mathrm{d}s^2 = -e^{2\Phi(r)}\mathrm{d}t^2+e^{2\Lambda(r)}\mathrm{d}r^2-2\omega(r)r^2\sin^2{\theta}\mathrm{d}t \mathrm{d}\phi+r^2\mathrm{d}\Omega^2,
\end{equation}
where $\mathrm{d}\Omega^2=\mathrm{d}\theta^2+\sin^2{\theta}\mathrm{d}\phi^2${, and $\Phi$ and $\Lambda$ are radial metric functions. We should state that neglecting higher-order terms in the Hartle-Thorne metric leads to the Schwarzchild metric with the frame-dragging correction. Hereafter we will name this metric the slow-rotation approximation of the Kerr metric and adopt the Boyer-Lindquist coordinate system ($t,r,\theta,\phi$). Despite being described by different background metrics, the linearised slow-rotation approximation of the metric for neutron stars and black holes is equivalent. The frame-dragging correction captures the effect of the differential rotation $\omega(r)$ which is associated with the angular velocity of a free-falling inertial frame \citep{Ravenhall1994}:}
\begin{equation}
    \omega(r)\equiv \frac{\mathrm{d}\phi}{\mathrm{d}t} = -\beta^{\phi}\approx 0.21\Omega_*\frac{R_s}{R_*-R_s}\left(\frac{R_*}{r}\right)^3,
\end{equation}
{where $R_s$ is the Schwarzschild radius.} 

This system is divided into two domains: the interior and the exterior of the star. The exterior part is well known, and the metric functions are given by
\begin{equation}
    e^{\Phi(r)} = e^{-\Lambda(r)} \equiv \alpha(r) = \sqrt{1-\frac{2M}{r}},~ \mathrm{for}~r>R_*,
\end{equation}
while the interior part is more complicated and requires knowledge of the constituents and structure of the star. Throughout this paper, we adopt a reduced model and consider a star with constant density, corresponding to the \textit{stiff-matter} equation of state \citep{Rezzolla2001a}.

\subsection{Maxwell's equations}

In this fixed background metric, Maxwell's equations take the form \citep{Komissarov2011}:
\begin{align}
    \boldsymbol{\nabla}\cdot\boldsymbol{E} &= 4\pi\rho,\label{eq4}\\
    \boldsymbol{\nabla}\cdot\boldsymbol{B} &= 0,\label{eq5}\\
    -\upartial_t\boldsymbol{B} &= \boldsymbol{\nabla}\times\left(\alpha\boldsymbol{E}+\boldsymbol{\beta}\times\boldsymbol{B}\right),\label{eq6}\\
    \upartial_t\boldsymbol{E} &= \boldsymbol{\nabla}\times\left(\alpha\boldsymbol{B}-\boldsymbol{\beta}\times\boldsymbol{E}\right)-\alpha\boldsymbol{j}+\rho\boldsymbol{\beta},\label{eq7}
\end{align}
where physical quantities $\rho$, $j$, $\boldsymbol{B}$ and $\boldsymbol{E}$ are measured by zero angular momentum observers (ZAMOs) in a frame that is corotating with absolute space. The shift vector, $\boldsymbol{\beta}$, corresponds to the relative motion between the absolute space and the spherical coordinate grid, i.e. ($r,\theta,\phi$). The lapse function, $\alpha$, is the ratio of the ticking rate of the local fiducial observer clock and the universal time $t$. In a sense, one can think of the lapse function as a converter of quantities measured with local clocks to the universal time coordinate. 

In this paper, we use the orthonormal basis vectors and vectorial components, i.e. $e_{\hat i} \equiv e_{i}/{h_{i}}$ and $A^{\hat i}\equiv {h_{i}} A^{i}$ such that $\boldsymbol{A} = A^{i}e_i = A^{\hat i}e_{\hat i}$, where $h_{i}^2$ are the diagonal terms of the spatial 3-metric. These basis vectors and vectorial components are often called the \textit{orthogonal basis} and \textit{physical components}, respectively, and are very useful as they allow three-dimensional vectorial operations to be easily generalised to curved geometry. For example, the curl of a generic vector is given by:
\begin{equation}
    \boldsymbol{\nabla}\times\boldsymbol{A}=\varepsilon^{ijk}\frac{\upartial A_j}{\upartial x^i}e_k = \frac{1}{h_1h_2h_3}
    \begin{vmatrix}
    h_1 e_{\hat{1}} & h_2 e_{\hat{2}} & h_3 e_{\hat{3}} \\ 
    \dfrac{\upartial}{\upartial{x^1}} & \dfrac{\upartial}{\upartial{x^2}} & \dfrac{\upartial}{\upartial{x^3}} \\ 
    h_1 A^{\hat{1}}  & h_2 A^{\hat{2}} & h_3 A^{\hat{3}}
    \end{vmatrix}.     
\end{equation}

Taking this into account, the general-relativistic Faraday \eqref{eq6} and Amp\`ere \eqref{eq7} equations in the exterior domain, component-wise, translate to:
\begin{align}
    r\sin{\theta}\upartial_tB^{\hat{r}}&=-\alpha\upartial_\theta\left(E^{\hat{\phi}}\sin{\theta}\right), \label{eq9}\\
    r\alpha^{-1}\upartial_tB^{\hat{\theta}}&=\upartial_r\left(\alpha r E^{\hat{\phi}}\right),\label{eq10}\\
    r\alpha^{-1}\upartial_tB^{\hat{\phi}}&=-\upartial_r\left(r\alpha E^{\hat\theta}\right)+\upartial_\theta E^{\hat r} + \sin{\theta}\upartial_r\left(\omega r^2B^{\hat r}\right)+\nonumber\\
    &+ \omega r \alpha^{-1}\upartial_\theta\left(\sin{\theta}B^{\hat\theta}\right),\label{eq11}\\
    r\sin{\theta}\upartial_tE^{\hat{r}}&=\alpha\upartial_\theta\left(B^{\hat{\phi}}\sin{\theta}\right)-\alpha r \sin{\theta}j^{\hat r},\label{eq12}\\
    r\alpha^{-1}\upartial_tE^{\hat{\theta}}&=-\upartial_r\left(\alpha r B^{\hat{\phi}}\right)- rj^{\hat \theta},\label{eq13}\\
    r\alpha^{-1}\upartial_tE^{\hat{\phi}}&=\upartial_r\left(r\alpha B^{\hat\theta}\right)-\upartial_\theta B^{\hat r}+\sin{\theta}\upartial_r\left(\omega r^2E^{\hat r}\right)+\nonumber\\
    &+\omega\alpha^{-1}r\upartial_\theta\left(\sin{\theta}E^{\hat\theta}\right)-rj^{\hat \phi}-\omega \alpha^{-1}r^2\sin{\theta}\rho,\label{eq14}
\end{align}
where we have neglected derivatives along the azimuthal direction, $\upartial_\phi$, as we will be tackling the aligned rotator case in an axisymmetric setup. In addition, we are interested in stationary solutions, meaning that $\upartial_t$ terms are set to zero from this point onwards.

\section{Stationary solutions to an aligned rotator}
\label{section:3}

In this section, we search for electromagnetic solutions of the general-relativistic Maxwell's equations presented in the previous section. We consider the aligned rotator case, where the magnetic and rotation axis are aligned ($\chi=0^\circ$). The aligned and the perpendicular ($\chi=90^\circ$) rotators constitute a basis to study the generic oblique rotator. However, in this paper, we solely focus on the first case. 

As was addressed in the literature \citep[e.g.][]{Petri2013}, the rotating dipole in a fixed background metric is a complex problem for which only approximate solutions were found. We will follow closely the approach first presented by \citet{Rezzolla2001a}, where each field component is expanded in powers of the frame-dragging frequency (or, Lense-Thirring frequency):
\begin{align}
    B^{\hat r} &= B^{\hat r}_{(0)_r}(r)\cos{\theta} + B^{\hat r}_{(1)}(r,\theta) + B^{\hat r}_{(2)}(r,\theta) + \mathcal{O}(\omega^3),\label{eq15}\\
    B^{\hat \theta} &= B^{\hat \theta}_{(0)_r}(r)\sin{\theta} + B^{\hat \theta}_{(1)}(r,\theta) + B^{\hat \theta}_{(2)}(r,\theta)+ \mathcal{O}(\omega^3),\\
    E^{\hat r} &= E^{\hat r}_{(0)}(r,\theta)+ E^{\hat r}_{(1)}(r,\theta) + E^{\hat r}_{(2)}(r,\theta)+ \mathcal{O}(\omega^3),\\
    E^{\hat \theta} &= E^{\hat \theta}_{(0)}(r,\theta) + E^{\hat \theta}_{(1)}(r,\theta) + E^{\hat \theta}_{(2)}(r,\theta)+ \mathcal{O}(\omega^3),\label{eq18}\\
    E^{\hat \phi} &= B^{\hat \phi} = 0,\label{eq19}
\end{align}
where the subscript in brackets gives the expansion order, e.g. $ B^{\hat r}_{(2)}\propto\omega^2$. Henceforth, we omit the explicit coordinate dependence and add it in subscript, e.g. $B^{\hat r}_{(0)_r}(r)\equiv B^{\hat r}_{(0)_r}$. The azimuthal components are zero due to the axisymmetric constraint. The magnetic field at zeroth-order has the dipolar angular dependence, but we allow the remaining radial and angular eigenfunctions to be self-consistently generated from Maxwell's equations. 

The approach we follow consists of inserting the ansatz \eqref{eq15}-\eqref{eq19} into equations \eqref{eq11} and \eqref{eq14}, and closing the system with the divergencelessness constraint for the magnetic field \eqref{eq5} and electric field in vacuum \eqref{eq4}. These equations are then expanded in powers of the frame-dragging frequency. In the following subsections, we will treat one order at a time, starting from the zeroth-order terms.

\subsection{Zeroth-order solutions}

The zeroth-order terms correspond to the case where we neglect the frame-dragging effect, i.e. a rotating neutron star in a fixed Schwarzschild background metric.

Following the steps detailed above, and looking for separable solutions of the field components, i.e. $E^{\hat r}_{(0)}(r,\theta)=E^{\hat r}_{(0)_r}E^{\hat r}_{(0)_\theta}$, we find:
\begin{align}
    \upartial_r\left(r\alpha E^{\hat \theta}_{(0)_r}\right)E^{\hat \theta}_{(0)_\theta} - E^{\hat r}_{(0)_r}\upartial_\theta \left(E^{\hat r}_{(0)_\theta}\right)&= 0,\label{eq20}\\
    \upartial_r\left(r\alpha B^{\hat \theta}_{(0)_r}\right)\sin\theta - B^{\hat r}_{(0)_r}\upartial_\theta \left(\cos\theta\right)&= 0,\label{eq21}\\
    \upartial_r\left(r^2 B^{\hat r}_{(0)_r}\right)\sin{\theta}\cos{\theta} + \alpha^{-1} r B^{\hat \theta}_{(0)_r}\upartial_\theta \left(\sin^2\theta\right)&= 0,\label{eq22}\\
    \upartial_r\left(r^2 E^{\hat r}_{(0)_r}\right)\sin{\theta}E^{\hat r}_{(0)_\theta} + \alpha^{-1} r E^{\hat \theta}_{(0)_r}\upartial_\theta \left(\sin\theta E^{\hat \theta}_{(0)_\theta}\right)&= 0,\label{eq23}
\end{align}
where equations \eqref{eq20} and \eqref{eq21} correspond to the collection of zeroth-order terms of equations \eqref{eq11} and \eqref{eq14}, respectively. Also, equations \eqref{eq22} and \eqref{eq23} are the collection of zeroth-order terms of equations \eqref{eq5} and \eqref{eq4}, respectively.

The angular dependence of the electric field components can be determined by decoupling the angular and radial parts of equations \eqref{eq20} and \eqref{eq23}. This condition can be achieved through:
\begin{align}
    E^{\hat \theta}_{(0)_\theta} &\propto \upartial_\theta \left(E^{\hat r}_{(0)_\theta}\right),\\
    \sin\theta E^{\hat r}_{(0)_\theta} &\propto \upartial_\theta \left(\sin\theta E^{\hat \theta}_{(0)_\theta}\right),
\end{align}
which leads to the angular eigenfunction solution
\begin{align}
    E^{\hat r}_{(0)_\theta} &= 3\cos^2\theta - 1,\\
    E^{\hat \theta}_{(0)_\theta} &= \sin\theta\cos\theta,
\end{align}
and, consequently,
\begin{align}
    \sin{\theta}\cos{\theta}\left[\upartial_r\left(r\alpha E^{\hat \theta}_{(0)_r}\right) + 6 E^{\hat r}_{(0)_r}\right]= 0&,\label{eq28}\\
    \sin{\theta}\left[\upartial_r\left(r\alpha B^{\hat \theta}_{(0)_r}\right)+B^{\hat r}_{(0)_r}\right]= 0&,\label{eq29}\\
    \sin{\theta}\cos{\theta}\left[\upartial_r\left(r^2 B^{\hat r}_{(0)_r}\right) + 2\alpha^{-1} r B^{\hat \theta}_{(0)_r}\right]= 0&,\label{eq30}\\
    \sin{\theta}\left(3\cos^2{\theta}-1\right)\left[\upartial_r\left(r^2E^{\hat r}_{(0)_r}\right)+\alpha^{-1}r E^{\hat \theta}_{(0)_r} \right]= 0&.\label{eq31}
\end{align}

The radial eigenfunctions can be determined by solving the following differential equations, obtained by combining equation \eqref{eq28} with \eqref{eq31} and equation \eqref{eq29} with \eqref{eq30}:
\begin{align}
    &\upartial_r\left(\alpha^2\upartial_r\left(r^2E^{\hat r}_{(0)_r}\right)\right)-6E^{\hat r}_{(0)_r}=0,\label{eq32}\\
    &\upartial_r\left(\alpha^2\upartial_r\left(r^2B^{\hat r}_{(0)_r}\right)\right)-2B^{\hat r}_{(0)_r}=0,\label{eq33}\\
    &B^{\hat \theta}_{(0)_r}=-\frac{\alpha}{2r}\upartial_r\left(r^2B^{\hat r}_{(0)_r}\right),\\
    &E^{\hat \theta}_{(0)_r}=-\frac{\alpha}{r}\upartial_r\left(r^2E^{\hat r}_{(0)_r}\right).
\end{align}

The solution to equations \eqref{eq32} and \eqref{eq33} can be obtained more easily if one recasts them into the Legendre form by introducing $x=1-r/M$. The final solution here presented satisfies the physical condition of the vanishing field amplitude at infinity:
\begin{align}
    B^{\hat r}_{(0)_r}&=\frac{\mathcal{C}_1}{8}\left[\ln\left(1-\frac{2M}{r}\right)+\frac{2M}{r}\left(1+\frac{M}{r}\right)\right],\label{eq36}\\
    B^{\hat \theta}_{(0)_r}&=-\frac{\alpha(r)\mathcal{C}_1}{8}\left[\ln\left(1-\frac{2M}{r}\right)+\frac{2M}{r}\left(1+\frac{M}{r-2M}\right)\right],\label{eq37}\\
    E^{\hat r}_{(0)_r}&=\frac{\mathcal{C}_2}{4}\left[\left(3-\frac{2r}{M}\right)\ln\left(1-\frac{2M}{r}\right)+\frac{2M}{3r}\left(3+\frac{M}{r}\right)-4\right],\label{eq38}\\
    E^{\hat \theta}_{(0)_r}&=-\frac{3\alpha(r)\mathcal{C}_2}{2}\left[\left(1-\frac{r}{M}\right)\ln\left(1-\frac{2M}{r}\right)-\frac{2M^2}{3r\left(r-2M\right)}-2\right],\label{eq39}
\end{align}
where $\mathcal{C}_1$ and $\mathcal{C}_2$ are integration constants found by applying the Newtonian limit and demanding the continuity of the solution across the star surface, see section~\ref{subsection:3.4}.

\subsection{First-order solutions}

In this section, we consider the inclusion of the first-order terms, i.e. the terms that are proportional to the frame-dragging frequency $\omega$. Following the same procedure as in the previous subsection, we obtain:
\begin{align}
    -&\upartial_r\left(r\alpha E^{\hat \theta}_{(1)}\right)+\upartial_\theta E^{\hat r}_{(1)}-3\omega r B^{\hat r}_{(0)_r}\sin\theta\cos\theta= 0,\label{eq40}\\
    &\upartial_r\left(r^2E^{\hat r}_{(1)}\right)\sin\theta+\alpha^{-1}r\upartial_\theta\left(\sin\theta~E^{\hat \theta}_{(1)}\right)= 0,\label{eq41}\\
    &\upartial_r\left(r\alpha B^{\hat \theta}_{(1)}\right)-\upartial_\theta B^{\hat r}_{(1)}-3\omega r E^{\hat r}_{(0)_r}\sin\theta\left(3\cos^2\theta-1\right)= 0,\label{eq42}\\
    &\upartial_r\left(r^2B^{\hat r}_{(1)}\right)\sin\theta+\alpha^{-1}r\upartial_\theta\left(\sin\theta~B^{\hat \theta}_{(1)}\right)= 0.\label{eq43}
\end{align}

From the equations above, we see that the first-order field components are more complicated as they depend explicitly on the radial eigenfunctions of the zeroth-order terms. Due to this increased complexity of including complicated source terms, we will address the electric and magnetic field components separately, starting with the electric field equations that are less involved.

\subsubsection{Electric field equations}

Once again, we look for separable solutions to the field components of the type: 
\begin{align}
    E^{\hat r}_{(1)}=E^{\hat r}_{(1)_r}E^{\hat r}_{(1)_\theta},\label{eq44}\\
    E^{\hat \theta}_{(1)}=E^{\hat \theta}_{(1)_r}E^{\hat \theta}_{(1)_\theta}.\label{eq45}
\end{align}

Inserting equations \eqref{eq44}-\eqref{eq45} into \eqref{eq40}-\eqref{eq41} yields the following decoupling conditions:
\begin{align}
    E^{\hat \theta}_{(1)_\theta}\propto\upartial_\theta E^{\hat r}_{(1)_\theta}\propto\sin\theta\cos\theta,\\
    E^{\hat r}_{(1)_\theta}\sin\theta\propto\upartial_\theta\left(\sin\theta~E^{\hat \theta}_{(1)_\theta}\right),
\end{align}

By imposing that $E^{\hat \theta}_{(1)_\theta}\equiv s_1\sin\theta\cos\theta$, it follows:
\begin{align}
    \upartial_\theta\left(\sin\theta~E^{\hat \theta}_{(1)_\theta}\right)=s_1\sin\theta\left(3\cos^2\theta-1\right),\\
    E^{\hat r}_{(1)_\theta}\sin\theta=s_1 s_2\sin\theta\left(3\cos^2\theta-1\right),\\
    \upartial_\theta E^{\hat r}_{(1)_\theta} = -6s_1s_2\sin\theta\cos\theta,
\end{align}
where $s_1$ and $s_2$ are constants of proportionality.

Selecting the proportionality constants such that the angular profiles depend only on explicit trigonometric expressions ($s_1=s_2=1$), yields:
\begin{align}
    \sin\theta\cos\theta\left[-\upartial_r\left(r\alpha E^{\hat \theta}_{(1)_r}\right)-6 E^{\hat r}_{(1)_r}-3\omega r B^{\hat r}_{(0)_r}\right]&= 0,\\
    \sin\theta\left(3\cos^2\theta-1\right)\left[\upartial_r\left(r^2 E^{\hat r}_{(1)_r}\right)+\alpha^{-1}r E^{\hat \theta}_{(1)_r}\right]&= 0,
\end{align}
and, hence,
\begin{align}
    &E^{\hat r}_{(1)_\theta}=\left(3\cos^2\theta-1\right),\\
    &E^{\hat \theta}_{(1)_\theta}=\sin\theta\cos\theta,\\
    &\upartial_r\left(\alpha^2\upartial_r\left(r^2 E^{\hat r}_{(1)_r}\right)\right)-6 E^{\hat r}_{(1)_r}-3\omega r B^{\hat r}_{(0)_r}= 0,\label{eq56}\\
    &E^{\hat \theta}_{(1)_r}=-\frac{\alpha}{r}\upartial\left(r^2 E^{\hat r}_{(1)_r}\right).\label{eq57}
\end{align}

The solutions to \eqref{eq56}-\eqref{eq57}, satisfying the physical condition of the vanishing field amplitude at infinity, are:
\begin{align}
    &E^{\hat r}_{(1)_r}=\mathcal{C}_3\left[\left(3-\frac{2r}{M}\right)\ln{\left(1-\frac{2r}{M}\right)}+\frac{2M}{r}\left(1+\frac{M}{3r}\right)-4\right]-\nonumber\\
    &\hspace{0.25cm}-\frac{2 \mathcal{C}_4}{3}\frac{M^2}{4r^2}\left[\ln{\left(1-\frac{2M}{r}\right)}+\frac{2M}{r}\right],\\
    &E^{\hat \theta}_{(1)_r}=-6\alpha(r)\mathcal{C}_3\left[\left(1-\frac{r}{M}\right)\ln{\left(1-\frac{2M}{r}\right)}-\frac{2M^2}{3r\left(r-2M\right)}-2\right]+\nonumber\\
    &\hspace{0.25cm}+\frac{2\alpha(r) \mathcal{C}_4}{3}\left[\frac{M^4}{r^3\left(r-2M\right)}\right],
\end{align}
where $\mathcal{C}_3$ and $\mathcal{C}_4=3\omega_0C_1/(8M^2)$ are integration constants determined later in this paper (see section \ref{subsection:3.4}), and $\omega_0$ is the frame-dragging frequency stripped of its radial dependence (i.e. $\omega(r)=\omega_0/r^3$).


\subsubsection{Magnetic field equations}

The solution that decouples the system of differential equations for the magnetic field components is:
\begin{align}
    B^{\hat r}_{(1)}&=B^{\hat r}_{(1)_{r1}}B^{\hat r}_{(1)_{\theta1}}+B^{\hat r}_{(1)_{r2}}B^{\hat r}_{(1)_{\theta2}},\label{eq60}\\
    B^{\hat \theta}_{(1)}&=B^{\hat \theta}_{(1)_{r1}}B^{\hat \theta}_{(1)_{\theta1}}+B^{\hat \theta}_{(1)_{r2}}B^{\hat \theta}_{(1)_{\theta2}}.\label{eq61}
\end{align}
Note that these solutions require an extra term compared to \eqref{eq44}-\eqref{eq45}, motivating the separate analysis from the electric field equations.

Inserting equations \eqref{eq60}-\eqref{eq61} into \eqref{eq42}-\eqref{eq43} yields the following decoupling conditions for the first component equations:
\begin{align}
    &B^{\hat \theta}_{(1)_{\theta1}}\propto\upartial_\theta B^{\hat r}_{(1)_{\theta1}}\propto\sin\theta\left(3\cos^2\theta-1\right),\\
    &B^{\hat r}_{(1)_{\theta1}}\sin\theta\propto\upartial_\theta\left(\sin\theta~B^{\hat \theta}_{(1)_{\theta1}}\right).
\end{align}

Proceeding in the same manner, we can now impose that $B^{\hat \theta}_{(1)_{\theta1}}\equiv s_1\sin\theta\left(3\cos^2\theta-1\right)$ and it follows:
\begin{align}
    &\upartial_\theta\left(\sin\theta~B^{\hat \theta}_{(1)_{\theta1}}\right)=4s_1\sin\theta\cos\theta\left(3\cos^2\theta-2\right),\\
    &B^{\hat r}_{(1)_{\theta1}}\sin\theta=4s_1s_2\sin\theta\cos\theta\left(3\cos^2\theta-2\right),\\
    &\upartial_\theta B^{\hat r}_{(1)_{\theta1}} = -12s_1s_2\sin\theta\left(3\cos^2\theta-1\right)-4s_1s_2\sin\theta.
\end{align}

We select the constants such that the angular eigenfunctions are composed of explicit trigonometric functions ($s_1=4s_2=1$):
\begin{align}
    &B^{\hat r}_{(1)_{\theta1}}=\cos\theta\left(3\cos^2\theta-2\right),\\
    &B^{\hat \theta}_{(1)_{\theta1}}=\sin\theta\left(3\cos^2\theta-1\right),\\
    &\sin\theta\left(3\cos^2-1\right)\left[\upartial_r\left(r\alpha B^{\hat \theta}_{(1)_{r1}}\right)+3 B^{\hat r}_{(1)_{r1}}-r\omega r E^{\hat r}_{(0)_{r}}\right] +\nonumber\\
    &\hspace{0.25cm}+\upartial_r\left(r\alpha B^{\hat \theta}_{(1)_{r2}}\right)B^{\hat \theta}_{(1)_{\theta2}}+B^{\hat r}_{(1)_{r1}}\sin\theta-B^{\hat r}_{(1)_{r2}}\upartial_\theta B^{\hat r}_{(1)_{\theta2}}=0,\\
    &\sin\theta\cos\theta\left(3\cos^2-2\right)\left[\upartial_r\left(r^2 B^{\hat r}_{(1)_{r1}}\right)+4\alpha^{-1}r B^{\hat \theta}_{(1)_{r1}}\right]+\nonumber\\
    &\hspace{0.25cm}\upartial_r\left(r^2 B^{\hat r}_{(1)_{r2}}\right)B^{\hat r}_{(1)_{\theta2}}\sin\theta+\alpha^{-1}r B^{\hat \theta}_{(1)_{r2}}\upartial_\theta\left(\sin\theta B^{\hat \theta}_{(1)_{\theta2}}\right)=0.
\end{align}

To decouple the second component equations it requires that:
\begin{align}
    &B^{\hat \theta}_{(1)_{\theta2}}\propto\upartial_\theta B^{\hat r}_{(1)_{\theta2}}\propto\sin\theta,\\
    &B^{\hat r}_{(1)_{\theta2}}\sin\theta\propto\upartial_\theta\left(\sin\theta~B^{\hat \theta}_{(1)_{\theta2}}\right).
\end{align}

Thus, the natural choice is $B^{\hat \theta}_{(1)_{\theta2}} = s_1\sin\theta$, which yields:
\begin{align}
    &\upartial_\theta\left(\sin\theta~B^{\hat \theta}_{(1)_{\theta2}}\right)=2s_1\sin\theta\cos\theta,\\
    &B^{\hat r}_{(1)_{\theta2}}\sin\theta=2s_1s_2\sin\theta\cos\theta,\\
    &\upartial_\theta B^{\hat r}_{(1)_{\theta2}} = -2s_1s_2\sin\theta.
\end{align}

For this case, the constants are $s_1=2s_2=1$, yielding:
\begin{align}
    &B^{\hat r}_{(1)_{\theta2}}=\cos\theta,\\
    &B^{\hat \theta}_{(1)_{\theta2}}=\sin\theta,\\
    &\sin\theta\left(3\cos^2-1\right)\left[\upartial_r\left(r\alpha B^{\hat \theta}_{(1)_{r1}}\right)+3 B^{\hat r}_{(1)_{r1}}-r\omega r E^{\hat r}_{(0)_{r}}\right] +\nonumber\\
    &\hspace{0.25cm}+\sin\theta\left[\upartial_r\left(r\alpha B^{\hat \theta}_{(1)_{r2}}\right)+B^{\hat r}_{(1)_{r1}}-B^{\hat r}_{(1)_{r2}}\right]=0,\\
    &\sin\theta\cos\theta\left(3\cos^2-2\right)\left[\upartial_r\left(r^2 B^{\hat r}_{(1)_{r1}}\right)+4\alpha^{-1}r B^{\hat \theta}_{(1)_{r1}}\right]+\nonumber\\
    &\hspace{0.25cm}+\sin\theta\cos\theta\left[\upartial_r\left(r^2 B^{\hat r}_{(1)_{r2}}\right)+2\alpha^{-1}r B^{\hat \theta}_{(1)_{r2}}\right]=0,
\end{align}
which gives rise to the radial differential equations for the magnetic field components:
\begin{align}
    &\upartial_r\left(\alpha^2\upartial_r\left(r^2 B^{\hat r}_{(1)_{r1}}\right)\right)-12 B^{\hat r}_{(1)_{r1}}+12\omega r E^{\hat r}_{(0)_{r}}=0,\\
    &\upartial_r\left(\alpha^2\upartial_r\left(r^2 B^{\hat r}_{(1)_{r2}}\right)\right)-2 B^{\hat r}_{(1)_{r2}}-2 B^{\hat r}_{(1)_{r1}}=0,\\
    &B^{\hat \theta}_{(1)_{r1}}=-\frac{\alpha}{4r}\upartial_r\left(r^2 B^{\hat r}_{(1)_{r1}}\right),\\
    &B^{\hat \theta}_{(1)_{r2}}=-\frac{\alpha}{2r}\upartial_r\left(r^2 B^{\hat r}_{(1)_{r2}}\right).
\end{align}

Each magnetic field component is comprised of two terms. Consequently, we have four decoupled differential equations. We highlight the fact that the second term, $B^{\hat r}_{(1)_{r2}}$, depends explicitly on the first one, $B^{\hat r}_{(1)_{r1}}$.

The final solution to the system of differential equations presented above, satisfying the vanishing field amplitude at infinity, is given by:
\begin{align}
    &B^{\hat r}_{(1)_{r1}}=\mathcal{C}_5\left[\frac{2M}{r}\left(6+\frac{M}{r}\right)+\frac{45r}{M}-75+\right.\nonumber\\
    &\hspace{0.25cm}+\left.\left(36+\frac{15r}{2M}\left(\frac{3r}{M}-8\right)\right)\ln{\left(1-\frac{2M}{r}\right)}\right]+\nonumber\\
    &\hspace{0.25cm}+\frac{\mathcal{C}_6}{6}\frac{M}{r}\left[\frac{2M}{r}\left(\frac{M}{r}-3\right)+\left(\frac{4M}{r}-3\right)\ln{\left(1-\frac{2M}{r}\right)}\right],\\
    &B^{\hat r}_{(1)_{r2}}=\frac{\mathcal{C}_7}{8}\left[\frac{2M}{r}\left(\frac{M}{r}+1\right)+\ln{\left(1-\frac{2M}{r}\right)}\right]+\nonumber \\
    &\hspace{0.25cm}+\mathcal{C}_5\left[\frac{2M}{r}\left(\frac{M}{r}+2\right)-15+\frac{9r}{M}+\frac{1}{2}\left(4-\frac{3r}{M}\right)^2\ln{\left(1-\frac{2M}{r}\right)}\right]-\nonumber\\
    &\hspace{0.25cm}-\frac{\mathcal{C}_6}{6}\left(\frac{M}{r}-2\right)\left[\left(\frac{M}{r}-1\right)\ln{\left(1-\frac{2M}{r}\right)}-\frac{2M}{r}\right],\\
    &B^{\hat \theta}_{(1)_{r1}}=-\alpha(r)\frac{M}{4r}\left\{6 \mathcal{C}_5\left[4-\frac{30r}{M}\left(1-\frac{r}{M}\right)+\frac{4M}{r-2M}+\right.\right.\nonumber\\
    &\hspace{0.25cm}+\left.\frac{3r}{M}\left(4-\frac{5r}{M}\left(2-\frac{r}{M}\right)\right)\ln{\left(1-\frac{2M}{r}\right)}\right]-\nonumber\\
    &\hspace{0.25cm}-\left.\frac{\mathcal{C}_6}{6}\left[\frac{2M}{r-2M}+\frac{2M}{r}\left(\frac{M}{r}+2\right)+3\ln{\left(1-\frac{2M}{r}\right)}\right]\right\},\\
    &B^{\hat \theta}_{(1)_{r2}}=-\alpha(r)\frac{M}{2r}\left\{\frac{\mathcal{C}_7}{4}\left[2+\frac{2M}{r-2M}+\frac{r}{M}\ln{\left(1-\frac{2M}{r}\right)}\right]\right.+\nonumber\\
    &\hspace{0.25cm}+2 \mathcal{C}_5\left[2\left(2-\frac{9r}{M}\left(1-\frac{r}{M}\right)+\frac{2M}{r-2M}\right)+\right.\nonumber\\
    &\hspace{0.25cm}+\left.\frac{r}{M}\left(\frac{3r}{M}-4\right)\left(\frac{3r}{M}-2\right)\ln{\left(1-\frac{2M}{r}\right)}\right]+\nonumber\\
    &\hspace{0.25cm}+\left.\frac{\mathcal{C}_6}{6}\left[\frac{M}{r}-8-\frac{3M}{r-2M}+\left(3-\frac{4r}{M}\right)\ln{\left(1-\frac{2M}{r}\right)}\right]\right\},
\end{align}
where $\mathcal{C}_5$, $\mathcal{C}_6=\omega_0C_2/M^2$ and $\mathcal{C}_7$ are constants of integration to be determined later on this paper, see section \ref{subsection:3.4}.

These terms constitute the frame-dragging correction to the magnetic field components and, to the extent of our knowledge, have not been described in previous works in analytical and closed form. An important aspect, here highlighted, is the fact that general relativity leads to even multipoles of the initial seeding field. In the current paper, we are seeding the dipolar magnetic field, and the outcome is the generation of a quadrupolar magnetic field correction. Despite not being shown here explicitly, the procedure presented can be used to determine the second-order correction terms:
\begin{align}
E^{\hat r}_{(2)_{\theta1}}&=\left(3\cos^2\theta -1\right),\\
E^{\hat r}_{(2)_{\theta2}}&=\left(15\cos^4\theta-15\cos^2\theta+2\right),\\
E^{\hat \theta}_{(2)_{\theta1}}&=\sin\theta\cos\theta,\\
E^{\hat \theta}_{(2)_{\theta2}}&=\sin\theta\cos\theta\left(3\cos^2\theta -2\right),\\
B^{\hat r}_{(2)_{\theta1}}&=\cos\theta\left(3\cos^2\theta-2\right),\\
B^{\hat r}_{(2)_{\theta2}}&=\cos\theta,\\
B^{\hat \theta}_{(2)_{\theta1}}&=\sin\theta\left(3\cos^2\theta-1\right),\\
B^{\hat \theta}_{(2)_{\theta2}}&=\sin\theta,
\end{align}
showing the appearance of an octopolar electric field component, and new corrections to previously existing multipolar amplitudes. However, these corrections are neglected in this paper as we are considering the slow-rotating approximation of the Kerr metric.

\subsection{Interior solution}

To obtain the complete solution one needs to know the interior solution inside the neutron star to get the integration coefficients via the interface matching conditions. In this subsection, we follow the same methodology as \citet{Rezzolla2001a}, assuming the neutron star as a perfect conductor ($\sigma\rightarrow\infty$) and looking for a radially uniform interior magnetic field solution $\left(\mathrm{i.e., }~B^{\hat{k}}(r,\theta)=B^{\hat{k}}(\theta)\right)$, corresponding to the \textit{stiff-matter} equation of state case.

\subsubsection{Perfect conductor compact neutron star}

The perfect conductor constraint provides a way to solve Maxwell's equations using the general-relativistic Ohm's law \citep[in the ZAMO frame, e.g.][]{Rezzolla2001a}: 
\begin{align}
    j^{\hat{t}}&=\rho + \sigma\frac{\bar\omega r \sin\theta}{e^{\Phi}}E^{\hat{\phi}}_{\mathrm{in}},\\
    j^{\hat{r}}&=\sigma\left(E^{\hat{r}}_{\mathrm{in}}-\frac{\bar\omega r \sin\theta}{e^{\Phi}}B^{\hat{\theta}}_{\mathrm{in}}\right),\\
    j^{\hat{\theta}}&=\sigma\left(E^{\hat{\theta}}_{\mathrm{in}}+\frac{\bar\omega r \sin\theta}{e^{\Phi}}B^{\hat{r}}_{\mathrm{in}}\right),\\
    j^{\hat{\phi}}&=\sigma E^{\hat{\phi}}_{\mathrm{in}}+\frac{\bar\omega r \sin\theta}{e^{\Phi}}\rho,\\
    \bar\omega &= \Omega_* - \omega,\label{eq101}
\end{align}
as it allows to write the electric field components as a function of the magnetic field:
\begin{align}
    E^{\hat{r}}_{\mathrm{in}}&=\frac{\bar\omega r \sin\theta}{e^{\Phi}}B^{\hat{\theta}}_{\mathrm{in}},\label{eq102}\\
    E^{\hat{\theta}}_{\mathrm{in}}&=-\frac{\bar\omega r \sin\theta}{e^{\Phi}}B^{\hat{r}}_{\mathrm{in}}.\label{eq103}
\end{align}

The interior solution is then fully determined by solving the interior magnetic field equations that can be obtained via equations \eqref{eq5} and the azimuthal component of \eqref{eq6}:
\begin{align}
    \sin{\theta}\upartial_r\left(r^2 B^{\hat r}_{\mathrm{in}}\right)&+e^{\Lambda}r\upartial_{\theta}\left(\sin{\theta} B^{\hat \theta}_{\mathrm{in}}\right) =0,\label{eq104}\\
    \upartial_r\left(r e^{\Phi} E^{\hat \theta}_{\mathrm{in}}\right) &- e^{\Phi+\Lambda}\upartial_\theta E^{\hat r}_{\mathrm{in}}-\nonumber\\
    &-\sin{\theta}\upartial_r\left(\omega r^2 B^{\hat r}_{\mathrm{in}}\right)-\omega e^{\Lambda}r \upartial_\theta\left(\sin\theta B^{\hat \theta}_{\mathrm{in}}\right)= 0\label{eq105},
\end{align}
where we have already applied the stationary condition, i.e. $\upartial_t B^{\hat \phi}=0$, and the axisymmetry condition, i.e. $\upartial_\phi B^{\hat \phi}=0$. In fact, equation \eqref{eq105} reduces to \eqref{eq104} by the inclusion of equations \eqref{eq101}-\eqref{eq103}. With the assumption of the radially uniform interior magnetic field solution, we obtain:
\begin{equation}
    2\sin{\theta}B^{\hat r}_{\mathrm{in}}+e^{\Lambda}\upartial_\theta\left(\sin\theta B^{\hat \theta}_{\mathrm{in}}\right)= 0,\label{eq106}
\end{equation}
that is valid for all expansion orders.

Also, for the perfect conductor~/~vacuum interface, the matching conditions are the continuity of the normal magnetic field component ($B^{\hat{r}}$) and transverse electric field components ($E^{\hat{\theta}}$ and $E^{\hat{\phi}}$):
\begin{align}
    B^{\hat{r}}_{\mathrm{in}}(r=R_*)&=B^{\hat{r}}_{\mathrm{out}}(r=R_*),\label{eq107}\\
    E^{\hat{\theta}}_{\mathrm{in}}(r=R_*)&=E^{\hat{\theta}}_{\mathrm{out}}(r=R_*),\label{eq108}\\
    E^{\hat{\phi}}_{\mathrm{in}}(r=R_*)&=E^{\hat{\phi}}_{\mathrm{out}}(r=R_*) = 0.
\end{align}

From equation \eqref{eq107}, it makes sense to look for an interior solution that has the same angular dependence as the exterior solution, yielding:
\begin{equation}
B^{\hat r}_{\mathrm{in}}(\theta)=\bar{B}^{\hat r}_{(0)_r}\cos{\theta}+\bar{B}^{\hat r}_{(1)_{r1}}\cos{\theta}\left(3\cos^2\theta-2\right)+\bar{B}^{\hat r}_{(1)_{r2}}\cos{\theta},\label{eq110}
\end{equation}
with $\bar{B}^{\hat r}_{(0)_r}$, $\bar{B}^{\hat r}_{(1)_{r1}}$ and $\bar{B}^{\hat r}_{(1)_{r2}}$ being constants. This rational allows the determination of the $B^{\hat\theta}$ components via equation \eqref{eq106}:
\begin{align}
B^{\hat \theta}_{\mathrm{in}}(\theta)=\bar{B}^{\hat \theta}_{(0)_r}\sin{\theta}+\bar{B}^{\hat \theta}_{(1)_{r1}}\sin{\theta}\left(3\cos^2\theta-1\right)+\bar{B}^{\hat \theta}_{(1)_{r2}}\sin{\theta},\label{eq111}
\end{align}
with
\begin{align}
\bar{B}^{\hat \theta}_{(0)_r}&=-\bar{B}^{\hat r}_{(0)_r} e^{-\Lambda},\label{eq112}\\
\bar{B}^{\hat \theta}_{(1)_{r1}}&=-\bar{B}^{\hat r}_{(1)_{r1}} e^{-\Lambda}/2,\label{eq113}\\
\bar{B}^{\hat \theta}_{(1)_{r2}}&=-\bar{B}^{\hat r}_{(1)_{r2}} e^{-\Lambda}.\label{eq114}
\end{align}

Equations \eqref{eq102}-\eqref{eq103} yield then:
\begin{align}
    E^{\hat{r}}_{\mathrm{in}}&=\frac{\bar\omega r \sin^2\theta}{e^{\Phi}}\left(\bar{B}^{\hat \theta}_{(0)_r}+\bar{B}^{\hat \theta}_{(1)_{r1}}\left(3\cos^2\theta-1\right)+\bar{B}^{\hat \theta}_{(1)_{r2}}\right),\label{eq115}\\
    E^{\hat{\theta}}_{\mathrm{in}}&=-\frac{\bar\omega r \sin\theta\cos\theta}{e^{\Phi}}\left(\bar{B}^{\hat r}_{(0)_r}+\bar{B}^{\hat r}_{(1)_{r1}}\left(3\cos^2\theta-2\right)+\bar{B}^{\hat r}_{(1)_{r2}}\right),\label{eq116}
\end{align}
which generalises the internal electric field solution found by \cite{Rezzolla2001a}. Also, equations \eqref{eq110}-\eqref{eq111} and \eqref{eq115}-\eqref{eq116} constitute the internal solution for the non-zero electromagnetic components that will now be matched to the exterior one.

\subsection{Determination of the constants of integration}
\label{subsection:3.4}

To obtain the complete electromagnetic exterior solution, one just needs to combine the matching conditions given by \eqref{eq107}-\eqref{eq108} with the appropriate Newtonian limits:
\begin{align}
    B^{\hat r}_{\mathrm{flat}}(r,\theta)=\lim_{M/r\rightarrow 0}B^{\hat r}(r,\theta)=\frac{2\mu}{r^3}\cos\theta,\label{eq117}\\
    B^{\hat \theta}_{\mathrm{flat}}(r,\theta)=\lim_{M/r\rightarrow 0}B^{\hat \theta}(r,\theta)=\frac{\mu}{r^3}\sin\theta,\label{eq118}
\end{align}
where $\mu$ is the dipolar moment of the neutron star. As before, we will determine the integration coefficients one expansion order at a time.

\subsubsection{Matched zeroth-order solution}

When applying the Newtonian limit, only the zeroth-order terms in equations \eqref{eq36}-\eqref{eq37} survive and are capable to construct the radial dependence of the limits in \eqref{eq117}-\eqref{eq118}:
\begin{align}
    \lim_{M/r\rightarrow0}B^{\hat r}_{(0)_{r}}(r)=\frac{2\mu}{r^3},\\
    \lim_{M/r\rightarrow0}B^{\hat \theta}_{(0)_{r}}(r)=\frac{\mu}{r^3},
\end{align}
which are satisfied when
\begin{equation}
    \mathcal{C}_1 = -\frac{6\mu}{M^3},\label{eq121}
\end{equation}
and, consequently,
\begin{align}
    B^{\hat r}_{(0)_{r}}(r)&=-\frac{3\mu}{4M^3}\left[\ln\left(1-\frac{2M}{r}\right)+\frac{2M}{r}\left(1+\frac{M}{r}\right)\right],\label{eq122}\\
    B^{\hat \theta}_{(0)_{r}}(r)&=\frac{3\mu}{4M^3}\sqrt{1-\frac{2M}{r}}\left[\ln\left(1-\frac{2M}{r}\right)+\frac{2M}{r}\left(1+\frac{M}{r-2M}\right)\right].\label{eq123}
\end{align}

In the absence of rotation, expressions \eqref{eq122}-\eqref{eq123} correspond to the only non-zero electromagnetic components of a static dipolar field in a Schwarzschild background metric and in the aligned rotator configuration. These expressions coincide with the ones originally found in the work by \citet{Ginzburg1964,Anderson1970} \citep[see also expressions (90)-(91) in][]{Rezzolla2001a}.

As for the interior magnetic field components, equation \eqref{eq107} forces:
\begin{equation}
    \bar{B}^{\hat r}_{(0)_r} = B^{\hat r}_{(0)_{r}}(r=R_*) = -\frac{3\mu}{4M^3}\left[\ln\left(1-\frac{2M}{R_*}\right)+\frac{2M}{R_*}\left(1+\frac{M}{R_*}\right)\right],
\end{equation}
which determines $\bar{B}^{\hat \theta}_{(0)_r}$ via equation \eqref{eq112}. This leads to the determination of the zeroth-order electric field components given by equations \eqref{eq102}-\eqref{eq103} (or, \eqref{eq115}-\eqref{eq116}), and magnetic field components via equations \eqref{eq110}-\eqref{eq111}:
\begin{align}
    E^{\hat{r}\;0^{th}}_{\mathrm{in}}&=\frac{\Omega_* r \sin^2\theta}{e^{\Phi}}\bar{B}^{\hat \theta}_{(0)_r},\\
    E^{\hat{\theta}\;0^{th}}_{\mathrm{in}}&=-\frac{\Omega_* r \sin\theta\cos\theta}{e^{\Phi}}\bar{B}^{\hat r}_{(0)_r},\\
    B^{\hat{r}\;0^{th}}_{\mathrm{in}}&=\bar{B}^{\hat r}_{(0)_r}\cos\theta,\\
    B^{\hat{\theta}\;0^{th}}_{\mathrm{in}}&=\bar{B}^{\hat \theta}_{(0)_r}\sin{\theta}.
\end{align}

Imposing the continuity condition on the $\theta$ component of the electric field (i.e., equation \eqref{eq108}), gives:
\begin{align}
\mathcal{C}_2=&\frac{2\Omega_* R_* \bar{B}^{\hat r}_{(0)_r}}{3\alpha^2(R_*)}\times\nonumber\\
&\times\left[\left(1-\frac{R_*}{M}\right)\ln\left(1-\frac{2M}{R_*}\right)-\frac{2M^2}{3R_*\left(R_*-2M\right)}-2\right]^{-1},\label{eq129}
\end{align}
which, in par with equation \eqref{eq121}, completes the determination of the integration constants of the zeroth-order exterior solution, presented in equations \eqref{eq36}-\eqref{eq39}.

As the zeroth-order terms do not account for the frame-dragging effect, then the solution found corresponds to the case of a rotating neutron star with a static Schwarzschild background metric. It is important to notice that this solution coincides with expressions (97)-(99) and (134)-(136) in \citet{Rezzolla2001a} for the case of an intrinsic aligned dipolar field. Also, the radial eigenfunctions for the electric field satisfy the Newtonian limit solution:
\begin{align}
\lim_{M/r\rightarrow0} E^{\hat r}_{(0)_{r}}(r) = - \frac{\mu\Omega_*R_*^2}{r^4},\\
\lim_{M/r\rightarrow0} E^{\hat \theta}_{(0)_{r}}(r) = - \frac{2\mu\Omega_*R_*^2}{r^4},
\end{align}
reducing to the aligned rotating magnetised neutron star solution in Minkowski background metric, found by \citet{Deutsch1955}.

\subsubsection{Matched first-order solution}

The approach to determine the first-order integration constants is very similar to the zeroth-order ones. We start by noticing that the integration constants from the source terms of the differential equations, i.e. $\mathcal{C}_4$ and $\mathcal{C}_6$, are already determined via equations \eqref{eq121} and \eqref{eq129}.

Imposing the matching condition \eqref{eq108} for the first-order terms of equations \eqref{eq116} and \eqref{eq45}, leads to:
\begin{align}
    -\frac{R_*}{\alpha(R_*)} \left( -\omega_0\bar{B}^{\hat r}_{(0)_r}+\Omega_*\bar{B}^{\hat r}_{(1)_{r1}} \left( 3\cos^2\theta-2 \right)+\Omega_*\bar{B}^{\hat r}_{(1)_{r2}} \right) = \nonumber \\
    = E^{\hat \theta}_{(1)_{r}}(R_*),\label{eq133}
\end{align}
where we have already simplified the angular dependence on both sides of the equation. However, as the term proportional to $\bar{B}^{\hat r}_{(1)_{r1}}$ still has an angular dependence, then this constant must be zero. This is, of course, a consequence of not allowing other multipoles for the interior solution. In addition, we will not allow self-generated magnetic field components within the star. This means that $\bar{B}^{\hat r}_{(1)_{r2}}$ must also vanish such that the seeded interior magnetic field given by equations \eqref{eq110}-\eqref{eq111} is solely the zeroth-order dipolar field. These conditions can be further relaxed in future works. From equation \eqref{eq133}, it follows:
\begin{align}
    \bar{B}^{\hat r}_{(1)_{r1}} &= \bar{B}^{\hat \theta}_{(1)_{r1}}=0,\label{eq134}\\
    \bar{B}^{\hat r}_{(1)_{r2}} &= \bar{B}^{\hat \theta}_{(1)_{r2}}=0,\label{eq135}\\
    E^{\hat \theta}_{(1)_{r}}(R_*) &= \frac{R_*\omega_0}{\alpha(R_*)}\bar{B}^{\hat r}_{(0)_r}.\label{eq136}
\end{align}

Hence, imposing the matching condition \eqref{eq107} for the first-order terms of equations \eqref{eq110} and \eqref{eq60}, leads to:
\begin{align}
\bar{B}^{\hat r}_{(1)_{r1}} &= {B}^{\hat r}_{(1)_{r1}}(R_*),\label{eq137}\\
\bar{B}^{\hat r}_{(1)_{r2}} &= {B}^{\hat r}_{(1)_{r2}}(R_*),\label{eq138}
\end{align}
which were obtained by direct analysis of their angular dependencies. With equation \eqref{eq134}, equation \eqref{eq137} can be used to write $\mathcal{C}_5$ as a function of $\mathcal{C}_6$, that was already determined, as:
\begin{align}
    \mathcal{C}_5 = &-\frac{\mathcal{C}_6}{6}\frac{M}{R_*}\left[\frac{2M}{R_*}\left(\frac{M}{R_*}-3\right)+\left(\frac{4M}{R_*}-3\right)\ln{\left(1-\frac{2M}{R_*}\right)}\right]\times\nonumber\\
    &\times\left[\frac{2M}{R_*}\left(6+\frac{M}{R_*}\right)+\frac{45R_*}{M}-75+\right.\nonumber\\
    &\hspace{0.75cm}+\left.\left(36+\frac{15R_*}{2M}\left(\frac{3R_*}{M}-8\right)\right)\ln{\left(1-\frac{2M}{R_*}\right)}\right]^{-1}.\label{eq139}
\end{align}

Also, equation \eqref{eq138} can be used to determine $\mathcal{C}_7$ with the help of equation \eqref{eq135}:
\begin{align}
    &\mathcal{C}_7 =8\left[\frac{2M}{R_*}\left(\frac{M}{R_*}+1\right)+\ln{\left(1-\frac{2M}{R_*}\right)}\right]^{-1}\times\nonumber \\
    &\times\left\{\frac{\mathcal{C}_6}{6}\left(\frac{M}{R_*}-2\right)\left[\left(\frac{M}{R_*}-1\right)\ln{\left(1-\frac{2M}{R_*}\right)}-\frac{2M}{R_*}\right]-\right.\nonumber\\
    &\left.-\mathcal{C}_5\left[\frac{2M}{R_*}\left(\frac{M}{R_*}+2\right)-15+\frac{9R_*}{M}+\frac{1}{2}\left(4-\frac{3R_*}{M}\right)^2\ln{\left(1-\frac{2M}{R_*}\right)}\right]\right\},\label{eq140}
\end{align}
and equation \eqref{eq136} can be used to determine $\mathcal{C}_3$:
\begin{align}
\mathcal{C}_3 =& -\frac{1}{6\alpha(R_*)}\times\nonumber\\
& \times\left[\left(1-\frac{R_*}{M}\right)\ln{\left(1-\frac{2M}{R_*}\right)}-\frac{2M^2}{3R_*\left(R_*-2M\right)}-2\right]^{-1}\times\nonumber\\
& \times \left[\frac{2\alpha(R_*)}{3}\mathcal{C}_4\left(\frac{M^4}{R_*^3\left(R_*-2M\right)}\right)+\frac{R_*\omega_0}{\alpha(R_*)}\bar{B}^{\hat r}_{(0)_r}\right].\label{eq141}
\end{align}

The integration coefficients in equations \eqref{eq139}-\eqref{eq141} fully determine the exterior electromagnetic solution of a neutron star up to first-order, given by:
\begin{align}
    B^{\hat r} &= \left(B^{\hat r}_{(0)_{r}}(r)+B^{\hat r}_{(1)_{r1}}(r)\left(3\cos^2\theta-2\right)+B^{\hat r}_{(1)_{r2}}(r)\right)\cos{\theta},\label{eq142}\\
    B^{\hat \theta} &= \left(B^{\hat \theta}_{(0)_{r}}(r)+B^{\hat \theta}_{(1)_{r1}}(r)\left(3\cos^2\theta-1\right)+B^{\hat \theta}_{(1)_{r2}}(r)\right)\sin{\theta},\label{eq143}\\
    E^{\hat r} &= \left(E^{\hat r}_{(0)_{r}}(r) + E^{\hat r}_{(1)_{r}}(r)\right)\left(3\cos^2\theta-1\right),\label{eq144}\\
    E^{\hat \theta} &= \left(E^{\hat \theta}_{(0)_{r}}(r) + E^{\hat \theta}_{(1)_{r}}(r)\right)\sin{\theta}\cos{\theta},\label{eq145}\\
    E^{\hat \phi} &= B^{\hat \phi} = 0,\label{eq146}
\end{align}
and the corresponding electromagnetic interior solution:
\begin{align}
    B^{\hat r}_{\mathrm{in}}&=\bar{B}^{\hat r}_{(0)_r}\cos{\theta},\label{eq147}\\
    B^{\hat \theta}_{\mathrm{in}}&=\bar{B}^{\hat \theta}_{(0)_r}\sin{\theta},\label{eq148}\\
    E^{\hat{r}}_{\mathrm{in}}&=\frac{\bar\omega r \sin^2\theta}{e^{\Phi}}\bar{B}^{\hat \theta}_{(0)_r},\label{eq149}\\
    E^{\hat{\theta}}_{\mathrm{in}}&=-\frac{\bar\omega r \sin\theta\cos\theta}{e^{\Phi}}\bar{B}^{\hat r}_{(0)_r}.\label{eq150}
\end{align}

The electric field and the interior electromagnetic fields match exactly the solution found by \citet{Rezzolla2001a} for the aligned rotator. The same solution was later derived in expressions (56)-(57d) in \citet{Petri2013}. It is important to highlight that the frame-dragging correction to the exterior magnetic field in expressions \eqref{eq142}-\eqref{eq143} constitute a novel set of analytical electromagnetic solutions. Numerical solutions for the dipolar and multipolar fields were obtained in \citet{Petri2013} and \citet{Petri2017}, respectively.

\subsection{Solution analysis}

\begin{figure}
	\includegraphics[width=\columnwidth]{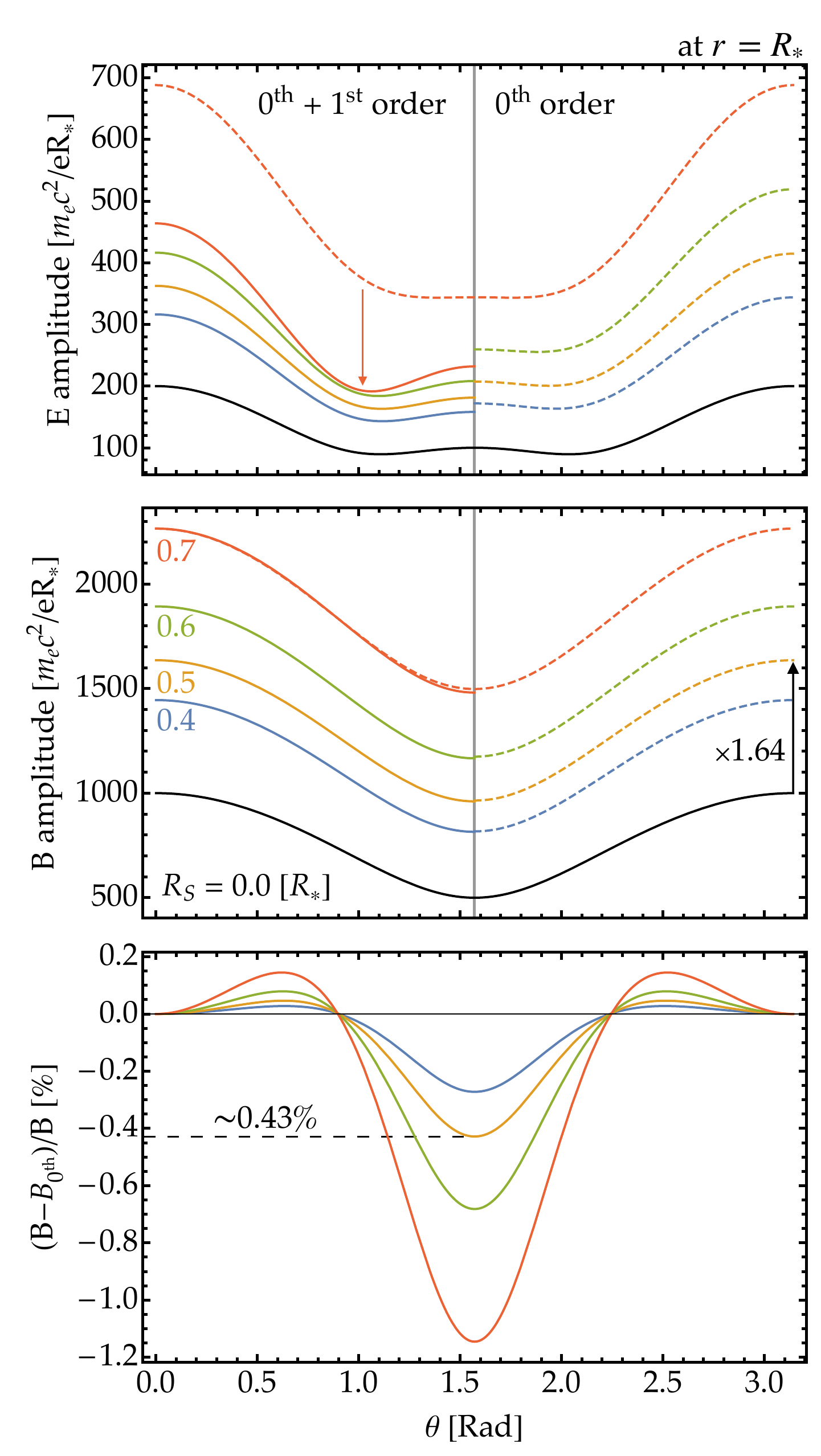}
    \caption{Electric and magnetic field amplitudes at the stellar surface, $r=R_*$, as a function of the polar angle $\theta$. The right side plots correspond to the zeroth-order fields, while the left side corresponds to the complete solution given in this paper. The bottom frame exhibits the correction percentage profile for the magnetic frame-dragging terms. Different colors correspond to different {values of compactness}. The Newtonian limit solution is presented in black.}
    \label{fig:E_B_amp}
\end{figure}

\begin{figure}
	\includegraphics[width=\columnwidth]{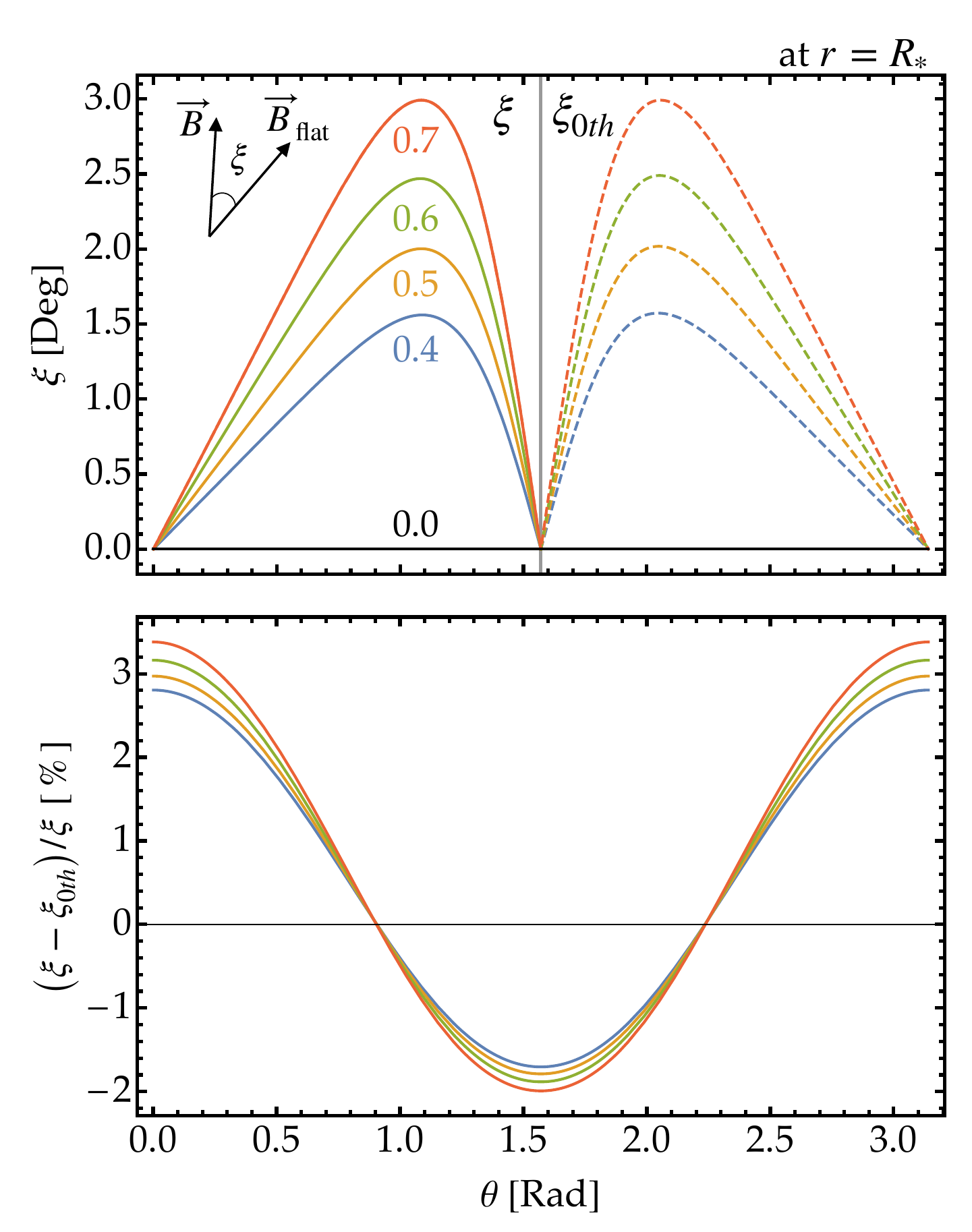}
    \caption{Vectorial angle, $\xi$, between the magnetic field in curved and flat spacetimes. In the top frame, it is shown how $\xi$ varies along $\theta$ with (left side) or without (right side) the magnetic frame-dragging correction. The bottom frame exhibits the correction percentage profile. Different colors correspond to different {values of compactness}. The Newtonian limit solution is presented in black.}
    \label{fig:xi_vec}
\end{figure}

It is important to notice that $\mathcal{C}_1$ is of order $\mu$, while $\mathcal{C}_2$ is of order $\mu\Omega_*$. This comes from the fact that the zeroth-order electric field is induced purely from the rotation of the star. In the same manner, $\mathcal{C}_4$ is of order $\mu\omega_0$, while $\mathcal{C}_6$ is of order $\mu\omega_0\Omega_*$. Just from this analysis, it is possible to establish the order of magnitude of the electromagnetic solution:
\begin{align}
  \mu > \mu\Omega_* &> \mu\omega_0 > \mu\omega_0\Omega_* > \hdots,\label{eq151}\\
  B_{(0)} > E_{(0)} &> E_{(1)} > \hspace{0.175cm}B_{(1)}\hspace{0.175cm} > \hdots,\label{eq152}
\end{align}
{if one recalls that $\omega_0 < \Omega_* \ll 1$.} This also states that the new terms, proportional to the frame-dragging frequency, are induced from the rotation of the background metric (electric field) or a nonlinear interplay between the stellar rotation and the metric rotation (magnetic field). Figure~\ref{fig:E_B_amp} shows the electric and magnetic field amplitudes for a normalised dipolar moment $\mu=500~[m_ec^2R_*^2/e]$ and a normalised stellar angular velocity $\Omega_*=0.2~[c~\mathrm{rad}~/R_*]$, realised for different {values of compactness}. This figure highlights the first conclusions by \citet{Ginzburg1964,Anderson1970} that the magnetic field amplitude is greatly enhanced when considering a curved spacetime geometry, at fixed dipolar moment. At the poles, the amplitude of the magnetic field could increase by a factor of $\sim 1.64$ for a typical compactness parameter $R_s \sim 0.5~[R_*]$. Figure~\ref{fig:E_B_amp} also exhibits the reduction of the induced electric field due to the frame-dragging effect as originally identified by \citet{Muslimov1986}. As for the frame-dragging correction to the magnetic field components, it shows that the new terms presented in this paper account for a $\sim0.43\%$ amplitude decrease in the equator for a typical compactness value. Figure~\ref{fig:xi_vec} shows that these corrections not only modify the magnetic amplitude but also influence their local vectorial properties. The angle $\xi$ between the flat spacetime magnetic field vector and the curved counterpart is shown to be rotated up to $\sim 2$ degrees for typical compactness values. In addition, Figure~\ref{fig:xi_vec} also demonstrates that the novel magnetic frame-dragging terms account for percent level corrections when compared to the zeroth-order curved magnetic field vectorial angle, i.e. without the frame-dragging correction.  


Another important detail is the existence of surface charges and currents. The surface charge distribution, $\sigma_s$, is supported by the discontinuity on the radial component of the electric field across the stellar surface \citep{Rezzolla2001a}:
\begin{align}
\sigma_s = \frac{1}{4\pi}\left(E^{\hat r}(R_*)-E^{\hat r}_{\mathrm{in}}(R_*)\right).
\end{align}

Similarly, the surface currents, $i^{\hat\theta}$ and $i^{\hat\phi}$, are supported by the discontinuity on the transverse magnetic field components across the stellar surface \citep{Rezzolla2001a}:
\begin{align}
i^{\hat\theta} &= \frac{c}{4\pi}\left(B^{\hat \phi}(R_*)-B^{\hat \phi}_{\mathrm{in}}(R_*)\right)=0,\\
i^{\hat\phi} &= \frac{c}{4\pi}\left(B^{\hat \theta}(R_*)-B^{\hat \theta}_{\mathrm{in}}(R_*)\right).
\end{align}

\begin{figure}
	\includegraphics[width=\columnwidth]{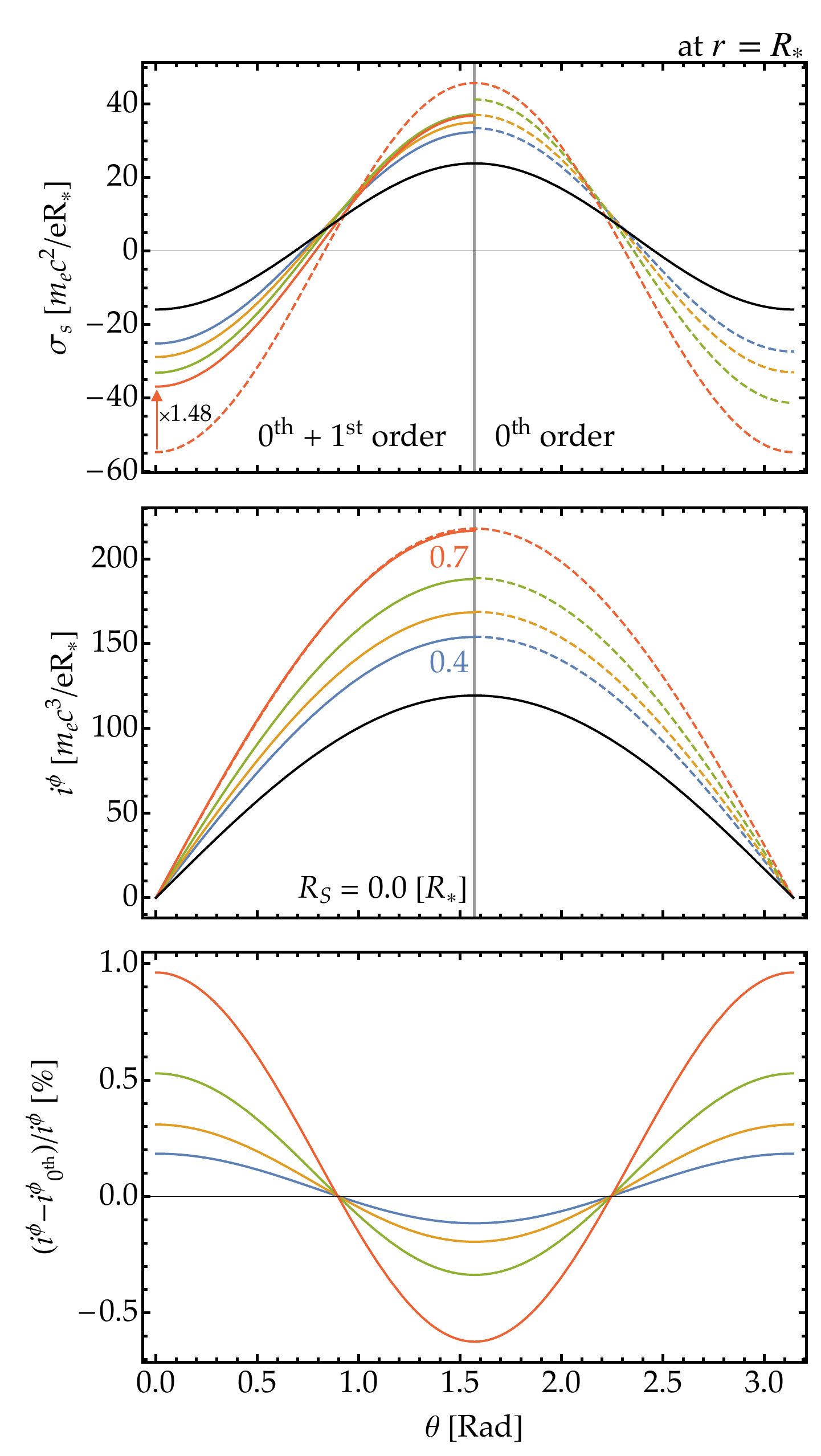}
    \caption{Surface charge and azimuthal current amplitudes as a function of the polar angle $\theta$ at the stellar surface. The right side plots correspond to the zeroth-order fields, while the left side corresponds to the complete solution given in this paper. The bottom frame exhibits the correction percentage profile for the surface current due to the magnetic frame-dragging terms. Different colors correspond to different {values of compactness}. The Newtonian limit solution is presented in black.}
    \label{fig:sigma_iphi_amp}
\end{figure}

Figure~\ref{fig:sigma_iphi_amp} shows the charge and azimuthal current amplitudes for the same numerical parameters. This figure shows that the frame-dragging correction does influence the surface charges by a factor of {$\sim1.14$} and the azimuthal surface current has a $\sim0.31\%$ amplitude decrease, both for a typical compactness value, and at the polar cap.

It should be stressed that these analytical results were obtained assuming that the interior magnetic field has a pure dipolar configuration. If multipoles were allowed inside the star, the interface conditions would allow non-trivial frame-dragging corrections at the stellar surface, which could lead to higher amplitude corrections.

\section{general-relativistic particle-in-cell code}
\label{section:4}
\begin{figure}
	\includegraphics[width=\columnwidth]{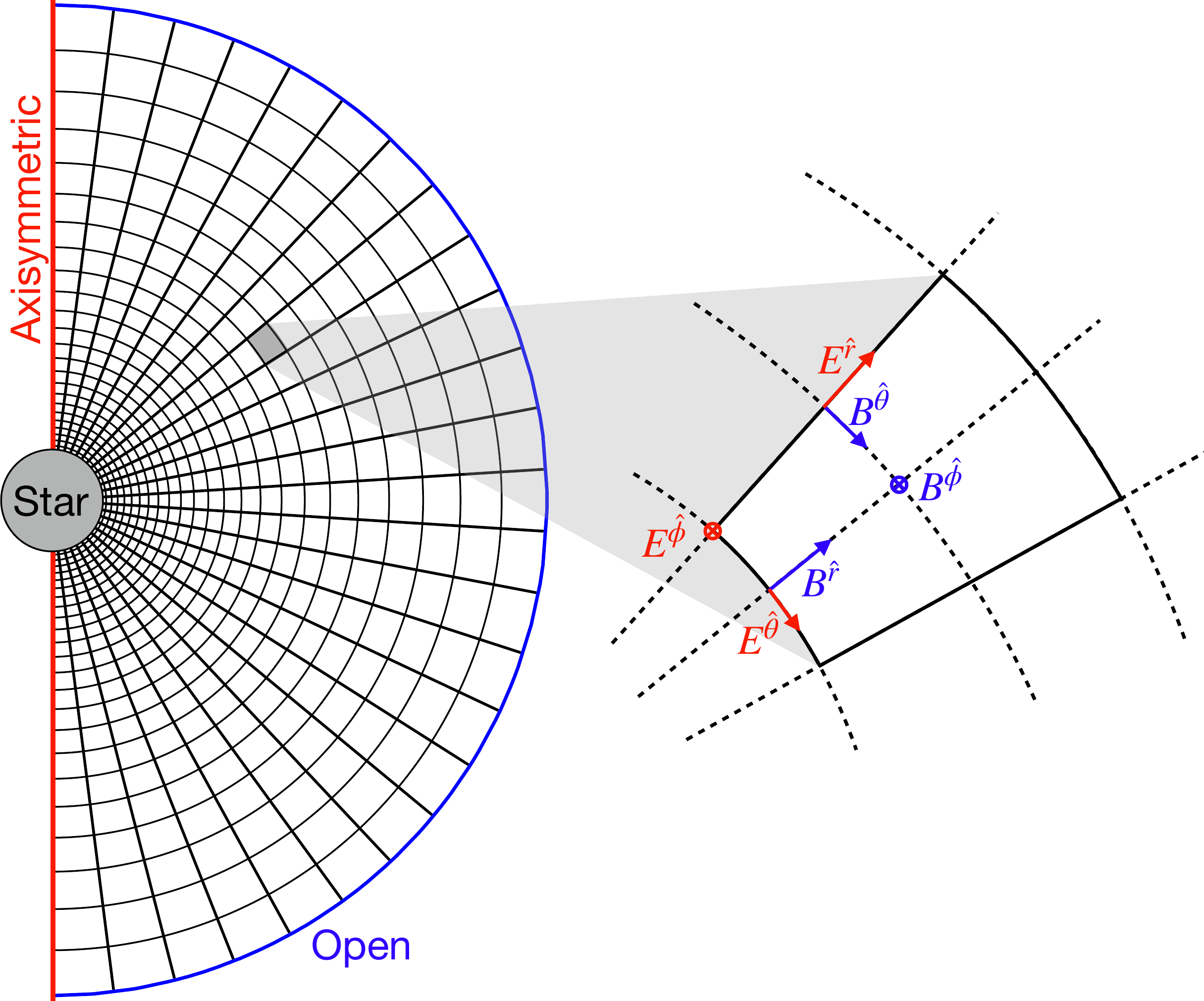}
    \caption{Spherical (r,$\theta$) grid with domain boundary conditions and electromagnetic field components layout.}
    \label{fig:gridcell}
\end{figure}

To simulate the exterior vacuum solution of a compact neutron star, we implemented a new module capable of capturing the general-relativistic effects within the OSIRIS particle-in-cell (PIC) code framework \citet{Osiris}. This new module makes use of the 3+1 formalism described in section~\ref{section:2}, performing all the calculations in Boyer-Lindquist coordinates in the slow-rotation limit of the Kerr metric.
A complete description of the particle-in-cell code OSIRIS-GR will be given in a future paper; here we present a brief overview of the numerical methods used. The field solver consists of a generalised version of the Yee algorithm \citep{Yee}, where the electric and magnetic field components are discretised on a spherical $(r,\theta)$ grid, as shown in Figure~\ref{fig:gridcell}. 
The grid is body-fitted to the shape of the neutron star. In this way, the interior radial boundary of the numerical domain corresponds to the stellar surface, placed at $r=R_*=1~[R_*]$. The outer boundary is placed at $r=r_{\mathrm{max}}=32~[R_*]$. As we are studying an axisymmetric setup, the domain corresponds to a poloidal cut with $\theta$ from $0$ to $\pi$, i.e. from north to south pole. In the radial direction, we chose a logarithmically spaced grid to have a higher resolution close to the stellar surface. In the meridian direction, we use a uniformly spaced grid. As for the electromagnetic boundaries, we implemented a Mur outer radial boundary that mimics an open boundary. For the polar axis, which corresponds to the meridional domain boundaries, we set the azimuthal field components to zero on the axis, i.e. $E^{\hat \phi}=B^{\hat \phi}=0$, and mirror the remaining angular components. For the inner radial boundary, we adopt the rotating conducting conditions by providing the interior solutions found in equations \eqref{eq147}-\eqref{eq150}. This means that we start our simulations with the star already in full rotation, i.e. angular velocity equal to $\Omega_*=0.125~[c~\mathrm{rad}~/R_*]$, and with dipolar moment $\mu=4000~[m_ec^2R_*^2/e]$.

The simulations presented in the next section are done with 2048 cells in both directions and with a compactness parameter $R_s = 0.5~[R_*]$ unless specified otherwise.

\section{Results}
\label{section:5}

\begin{figure}
	\includegraphics[width=\columnwidth]{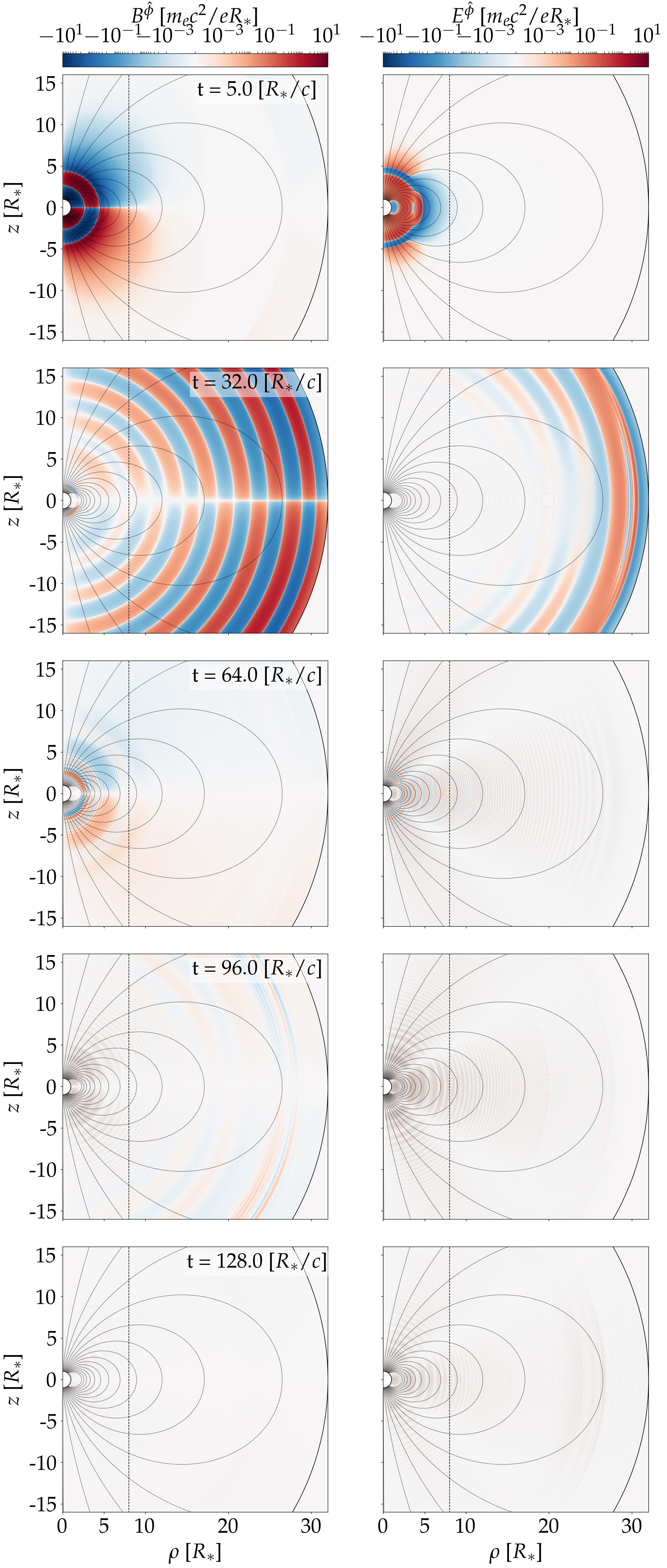}
    \caption{Temporal evolution of the azimuthal magnetic and electric field components for the \citet{Ginzburg1964} solution extended to rotation. The dashed black line represents the light cylinder distance and the poloidal field lines are represented in grey.}
    \label{fig:simulation_noise}
\end{figure}

In order to test the validity of both the code and the newly found solution, the exterior solution given by equations \eqref{eq142}-\eqref{eq146} will be initialized and the amplitude of the formed transient fields in the azimuthal field components will be examined. Ideally, if the exact solution to the system of equations was initialized, no transient would exist and these components would be null as in equation \eqref{eq146}. We start by initializing the solution found by \citet{Ginzburg1964} consisting of the zeroth-order terms of the magnetic field. This solution does not account for the frame-dragging effect but still captures the effect of the Schwarzschild fixed background metric. As we are interested in rotating neutron star solutions, we provide the zeroth-order electric field as well, despite, historically, not being the solution presented by \citet{Ginzburg1964}. Figure~\ref{fig:simulation_noise} shows how the transient propagates through the entire domain, being launched from the stellar surface and bouncing back and forth between the stellar surface and the outer radial boundary. Although the outer radial boundary is open-like, it is not a perfect absorbing layer and reflects the incoming wave with a much lower amplitude. It is important to note that after two complete bounces (i.e. after four light crossing times of the domain), the amplitude of these waves is negligible as seen for the last time shown in Figure~\ref{fig:simulation_noise}. We believe our boundary condition is still a better option than the standardised outer damping layer \citep[e.g.][]{Belyaev2015,Cerutti2015} as this layer also introduces waves into the system due to the damping of the background magnetic field, and modifies the amplitude of all the electromagnetic components close to it.

As mentioned before, the closer the initialized solution is to the exact solution the smaller the amplitude of the excited transient. In this sense, we can use the transient state as a probe for the three solutions present in this paper: (a) the solution found by \citet{Ginzburg1964} extended to account for the neutron star rotation as described above; (b) the solution found by \citet{Rezzolla2001a} that considers the frame-dragging correction to the electric field; (c) our solution that considers the frame-dragging correction in both the electric and magnetic fields. Recalling equation \eqref{eq152}, one sees that we are drawing closer to the exact solution as we are capturing {both first-order corrections}, hence, we expect the transient amplitude to decrease going from the first case to the last.
\begin{figure}
	\includegraphics[width=\columnwidth]{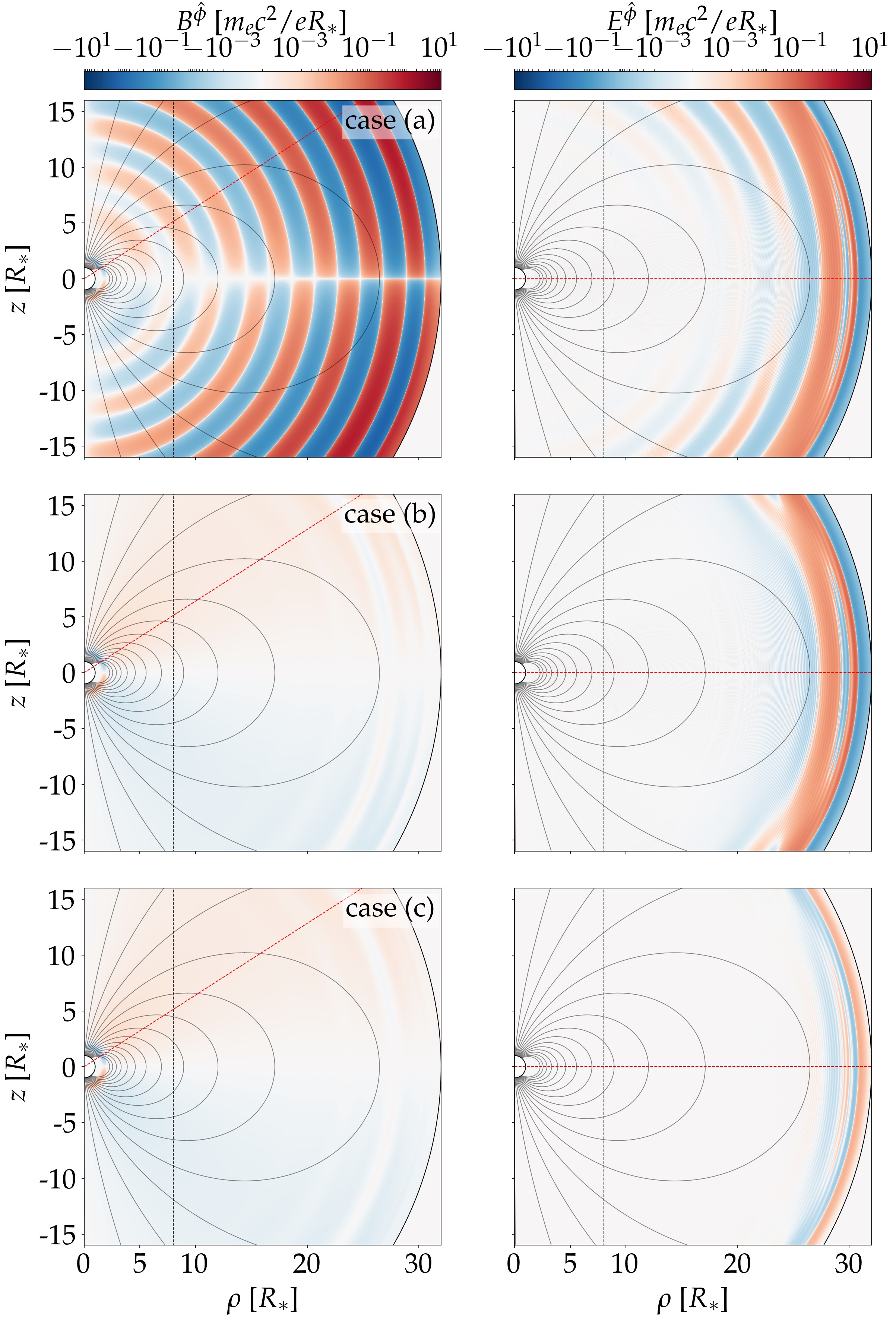}
    \caption{Comparison of the azimuthal transient amplitudes for the three solutions at $t=32~[R_*/c]$. The red dashed line represents the lineout locations for Figures~\ref{fig:bphi_ephi_lineouts} and \ref{fig:bphi_ephi_lineouts_zoom}.}
    \label{fig:bphi_ephi_comp}
\end{figure}
Figure~\ref{fig:bphi_ephi_comp} demonstrates this by showing that, indeed, the transient field decreases with the inclusion of {the first-order terms}. In {particular}, it shows that the inclusion of the frame-dragging correction to the electric (magnetic) field significantly reduces the azimuthal magnetic (electrical) field transient amplitude.
\begin{figure}
	\includegraphics[width=\columnwidth]{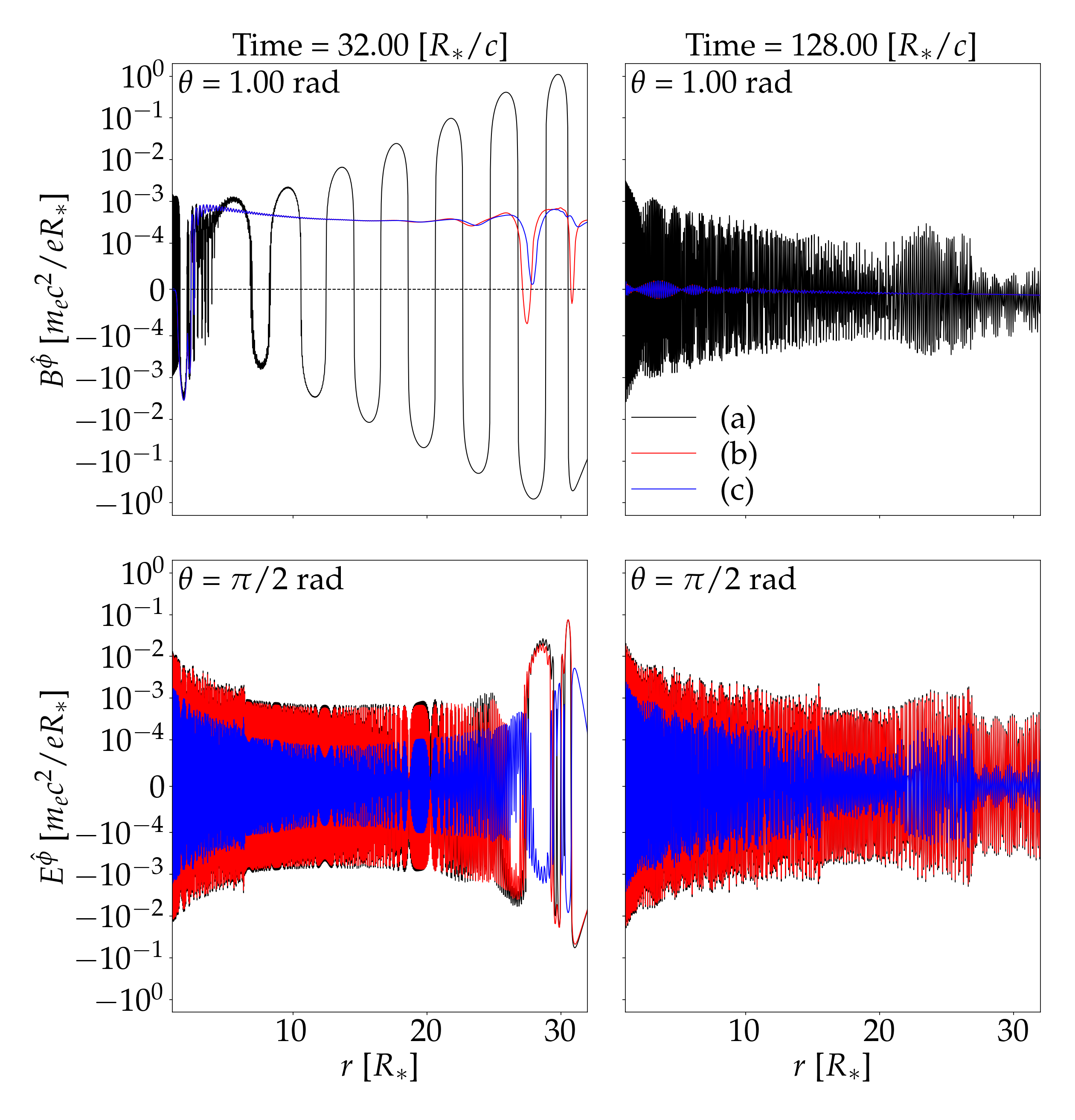}
    \caption{Transient amplitude at two times: $t=32$ and $128~[R_*/c]$. The upper panels show the azimuthal magnetic field for a cut at $\theta=1$~rad and the lower panels show the azimuthal electric field for a cut at the equator. The three solutions are given by black, red and blue colors, following the alphabetic order respectively.}
    \label{fig:bphi_ephi_lineouts}
\end{figure}
To better visualise this feature, we have taken a radial cut at $\theta=1.0~(\pi/2)~[\mathrm{rad}]$ for the azimuthal magnetic (electric) field component, as seen in the red dashed line in Figure~\ref{fig:bphi_ephi_comp}. The resulting lineouts are shown in Figure~\ref{fig:bphi_ephi_lineouts} for two distinct simulation times: $t=32.00$ (early-stage) and $128.00~[R_*/c]$ (late-stage). It is clear that our solution, case (c), has much smaller early and late-stage field amplitudes. It reduces the late-stage azimuthal electric field amplitude by approximately one order of magnitude, which means that the simulations are more accurate and stable at the expense of a more complicated initialization. Also, we demonstrate that for the stellar parameters chosen, solution (a) does not describe the late-stage very accurately, as expected. 

\begin{figure}
	\includegraphics[width=\columnwidth]{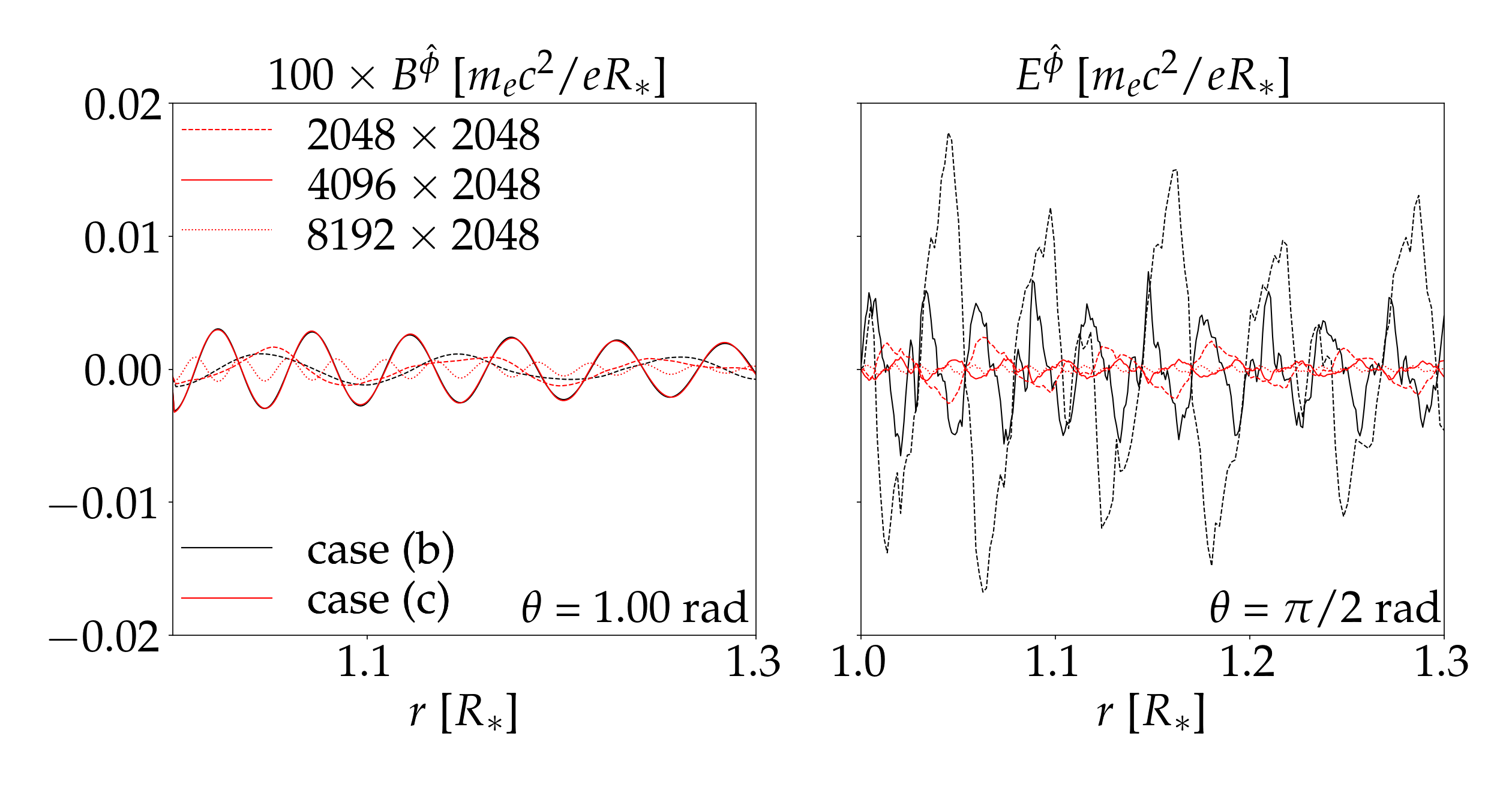}
    \caption{Close-up zoom of the stellar surface transient amplitude at $t=128~[R_*/c]$. The black and red lines correspond to the solution (b) and (c), respectively. Different radial resolutions correspond to different line styles: dashed, full and dotted correspond to $2048$, $4096$ and $8192$ grid cells, respectively. The azimuthal magnetic field amplitude was multiplied by 100 for better visualization.}
    \label{fig:bphi_ephi_lineouts_zoom}
\end{figure}

If we zoom in closer to the stellar surface for the late-stage lineouts, shown in Figure~\ref{fig:bphi_ephi_lineouts_zoom}, we can now analyse the grid resolution effect. Restricting ourselves to solutions (b) and (c), we can conclude that the azimuthal magnetic field component does not change significantly. However, if we look into the azimuthal electric field, we see that the amplitude of the late-stage component is reduced by half when the resolution is increased twofold. Interestingly, when using our solution at the lowest resolution, the late-stage amplitude of the transient is still smaller than the higher resolution run presented for solution (b), showing a higher stabilization of the obtained numerical solution and the potential to reduce the computational cost associated with this kind of studies. For this specific case, it would correspond to a speed-up of a factor of four.

\section{Conclusions}
\label{section:6}

Neutron stars comprise a set of compact objects where general-relativistic effects are relevant. This paper presents the solution of the magnetospheric electromagnetic fields to a massive neutron star in a vacuum background with an intrinsic dipolar magnetic moment. We summarise the analytic solutions obtained for an aligned rotator with infinite conductivity and extend them to consider the magnetic frame-dragging correction. Several equations of state models \citep[e.g.][]{Hebeler2013} predict that neutron stars can achieve compactness values up to $R_s \sim 0.6 [R_*]$. Therefore, we considered a typical value of $R_s \sim 0.5 [R_*]$ for the analysis of the derived solution. We show that the new terms account for a $0.43\%$ decrease in magnetic field strength at the equator and an average $1\%$ vectorial angle correction compared to previous solutions available in the literature. This solution modifies the external magnetic field configuration, leading to a self-consistent redistribution of the superficial azimuthal current.

We developed a new module for the OSIRIS particle-in-cell code capable of simulating the exterior magnetospheric problem of neutron stars with general relativity effects. This module performs all the calculations in Boyer-Lindquist coordinates in the slow-rotation limit of the Kerr metric. By prescribing the derived analytic solution to the exterior domain as an initial value problem, it is possible to compare its numerical stability with other solutions in the literature. Theoretically, both azimuthal components of the electric and magnetic fields should be zero in the exterior domain. In the early-stages of the simulation, the transient field launched can be used to probe the proximity between the prescribed solution and the exact one. We showed that the inclusion of the magnetic frame-dragging correction led to a significant reduction of the transient amplitude in both field components. In the late-stages of the simulation, the numerical solution converges to an oscillating reminiscent field whose amplitude is most affected by the initial prescribed solution. This was verified by noticing that the lowest resolution simulation with the electromagnetic frame-dragging correction has a lower reminiscent field amplitude than the highest resolution simulation considering only the electric field correction. Thus, demonstrating that simulations are more accurate and stable at the expense of a more complicated initialization. In particular, this corresponds to a reduction of the simulation runtime by a factor of four. The solution presented in this paper may be a handy tool to benchmark other particle-in-cell codes that rely on analytic solutions for the electromagnetic field initialization.

Future works could extend the present solution to include multipolar field contributions and a misalignment between the magnetic moment and stellar spin axis to approximate it from more realistic profiles; this would be a generalization of the approach presented in this work.

\section*{Acknowledgements}

This work is partially supported by the European Research Council (ERC-2015-AdG Grant 695088) {and FCT (Portugal) - Foundation for Science and Technology under the Project X-MASER No. 2022.02230.PTDC}. RT is supported by FCT (Portugal) (Grant PD/BD/142971/2018) in the framework of the Advanced Program in Plasma Science and Engineering (APPLAuSE, FCT Grant PD/00505/2018). The authors acknowledge useful discussions with Pablo J. Bilbao. {The authors thank the anonymous referee for the thoughtful and detailed comments that have significantly improved the manuscript}. All simulations presented were performed at LUMI within the EuroHPC-JU project EHPC-REG-2021R0038.

\section*{Data Availability}

Any data generated for or included in this article can be made available upon a reasonable request. All analytical solutions were obtained with a \texttt{Wolfram Mathematica 12} notebook. 
  



\bibliographystyle{mnras}
\bibliography{main} 








\bsp	
\label{lastpage}
\end{document}